\documentclass[fleqn,10pt]{wlscirep}
\usepackage[utf8]{inputenc}
\usepackage[T1]{fontenc}
\usepackage{hyphenat}
\usepackage{subcaption}

\title{Global gender differences in Wikipedia readership}

\author[1]{Isaac Johnson}
\author[2]{Florian Lemmerich}
\author[1]{Diego S\'aez-Trumper}
\author[3]{Robert West}
\author[2,4]{Markus Strohmaier}
\author[1]{Leila Zia}

\affil[1]{Wikimedia Foundation; first-name@wikimedia.org}
\affil[2]{RWTH Aachen University; first-name.last-name@cssh.rwth-aachen.de}
\affil[3]{EPFL; first-name.last-name@epf\/l.ch}
\affil[4]{GESIS - Leibniz Institute for the Social Sciences; first-name.last-name@gesis.org}

\begin{abstract}
 Wikipedia represents the largest and most popular source of encyclopedic knowledge in the world today, aiming to provide equal access to information worldwide. From a global online survey of 65,031 readers of Wikipedia and their corresponding reading logs, we present novel evidence of gender differences in Wikipedia readership and how they manifest in records of user behavior. More specifically we report that (1) women are underrepresented among readers of Wikipedia, (2) women view fewer pages per reading session than men do, (3) men and women visit Wikipedia for similar reasons, and (4) men and women exhibit specific topical preferences. Our findings lay the foundation for identifying pathways toward knowledge equity in the usage of online encyclopedic knowledge.
\end{abstract}

\begin{document}

\flushbottom
\maketitle
\thispagestyle{empty}

Equal access to encyclopedic knowledge represents a critical prerequisite for promoting equality more broadly in society and for an open and informed public at large. With almost 54 million articles written by roughly 500,000 monthly editors across more than 160 actively edited languages, Wikipedia is the most important source of encyclopedic knowledge for a global audience, and one of the most important knowledge resources available on the internet. Every month, Wikipedia attracts users on more than 1.5 billion unique devices from across the globe, for a total of more than 15 billion monthly pageviews.~\cite{stats2} Data about who is represented in this readership provides unique insight into the accessibility of encyclopedic knowledge worldwide. Unlike the well-documented gender gap amongst Wikipedia contributors~\cite{hill2013wikipedia,ford2017anyone,antin2011gender,lam2011wp,collier2012conflict,sichler2014gender}, however, little is known about the demographics of the people behind Wikipedia pageviews at a global scale and the similarities and differences between different populations of people in their readership.

Building on past research \cite{glott2010wikipedia,pew2011wikipedia,protonotarios2016similar,kim2014individual,lim2010gender,hinnosaar2019gender,selwyn2016students} and surveys\footnote{\url{https://meta.wikimedia.org/wiki/Category:Reader_surveys}} that have focused on specific reader populations, we provide insights into Wikipedia's global readership and their reading behavior through 16 large-scale surveys of more than 65,031 Wikipedia readers across 14 different language editions. We additionally link the readers' responses with the records\footnote{\url{https://wikitech.wikimedia.org/wiki/Analytics/Data_Lake/Traffic/Webrequest\#Current_Schema}} of their reading behavior on Wikipedia. Through this combination of survey responses and reading logs, we present novel evidence of gender differences in consuming Wikipedia and how they manifest in records of user behavior. In detail, we report that:

\noindent
\textit{Women are underrepresented among readers of Wikipedia.} Across 16 survey populations, men represent approximately two-thirds of Wikipedia readers on any given day. These results indicate that globally, substantial barriers still exist for women\footnote{While our surveys allowed readers to self-describe their gender identity, we only received sufficient data from men and women to conduct robust analyses. See the Methods for more details.} reading Wikipedia. A more balanced inclusion of women in Wikipedia's readership is demanded by Wikipedia's vision of imagining a world in which every human being can share in the sum of all knowledge\footnote{\url{https://wikimediafoundation.org/about/vision/}} as well as the need for \emph{knowledge equity}, identified as part of the Wikimedia Movement's 2030 strategic direction~\cite{wmstrategy}. Furthermore, a more balanced representation of gender identities among Wikipedia readers is a crucial prerequisite to achieving the diverse editor base necessary to expand Wikipedia's coverage to more languages and topics, as past research has established that editorship inequalities originate in part from inequalities in readership and awareness of the platform~\cite{shaw2018pipeline,khanna_readers_2011}.

\noindent
\textit{Women view fewer pages per reading session than men do.} We estimate that men generate around 72\% of pageviews in the surveyed Wikipedia editions on any given day, and many of the top-read articles on Wikipedia draw almost exclusively readers who are men. This finding has major implications on Wikipedia and on the downstream usage of Wikipedia's pageview data. It is a clear indicator that popularity-based recommendations and rankings on the platform that do not take into account the gender imbalance in readership~\cite{wulczyn2016growing} can further exacerbate the gender imbalances on Wikipedia. It is also important to note that Wikipedia's pageview data is used heavily by researchers~\cite{park2019estimating,smith2017using}, practitioners\cite{billboard}, and even policymakers\cite{bigdataPilot} for applications well beyond Wikipedia. Such usage of Wikipedia in the future should be viewed in the context of the findings of this study and account for the imbalances in the gender identities of readers behind Wikipedia pageviews, cf.~\cite{medin2017systems}.

\noindent
\textit{Men and women visit Wikipedia for similar reasons.} While men and women visit Wikipedia at dramatically different rates, the depth of knowledge that they seek, referred to as \emph{information need} for the remainder of this paper, and their triggers for reading Wikipedia, referred to as \emph{motivations}, are remarkably consistent across language editions.

\noindent
\textit{Men and women exhibit specific topical preferences.} Finally, across different language editions, we find that preferences vary by gender for certain topics. Interest in articles about sports, games, and mathematics is skewed towards men, while interest in articles about broadcasting, medicine, and entertainment is skewed towards women. We further observe \textit{self-focus bias}, such that women tend to read relatively more biographies of women than men do, whereas men tend to read relatively more biographies of men than women do. On Wikipedia, this finding mirrors the self-focus bias identified among contributors~\cite{hecht2010tower} and underlines the importance of the work conducted by Wikipedia volunteers through projects such as \emph{WikiProject Women in Red}\footnote{An example of such an effort on English Wikipedia: \url{https://en.wikipedia.org/wiki/Wikipedia:WikiProject_Women_in_Red}} to increase content about women in Wikipedia.

\section*{Results}

\subsection*{Women are underrepresented among readers of Wikipedia}
\begin{figure*}[ht]
    \centering
    \begin{subfigure}[t]{0.5\textwidth}
        \centering
        \includegraphics[width=\textwidth]{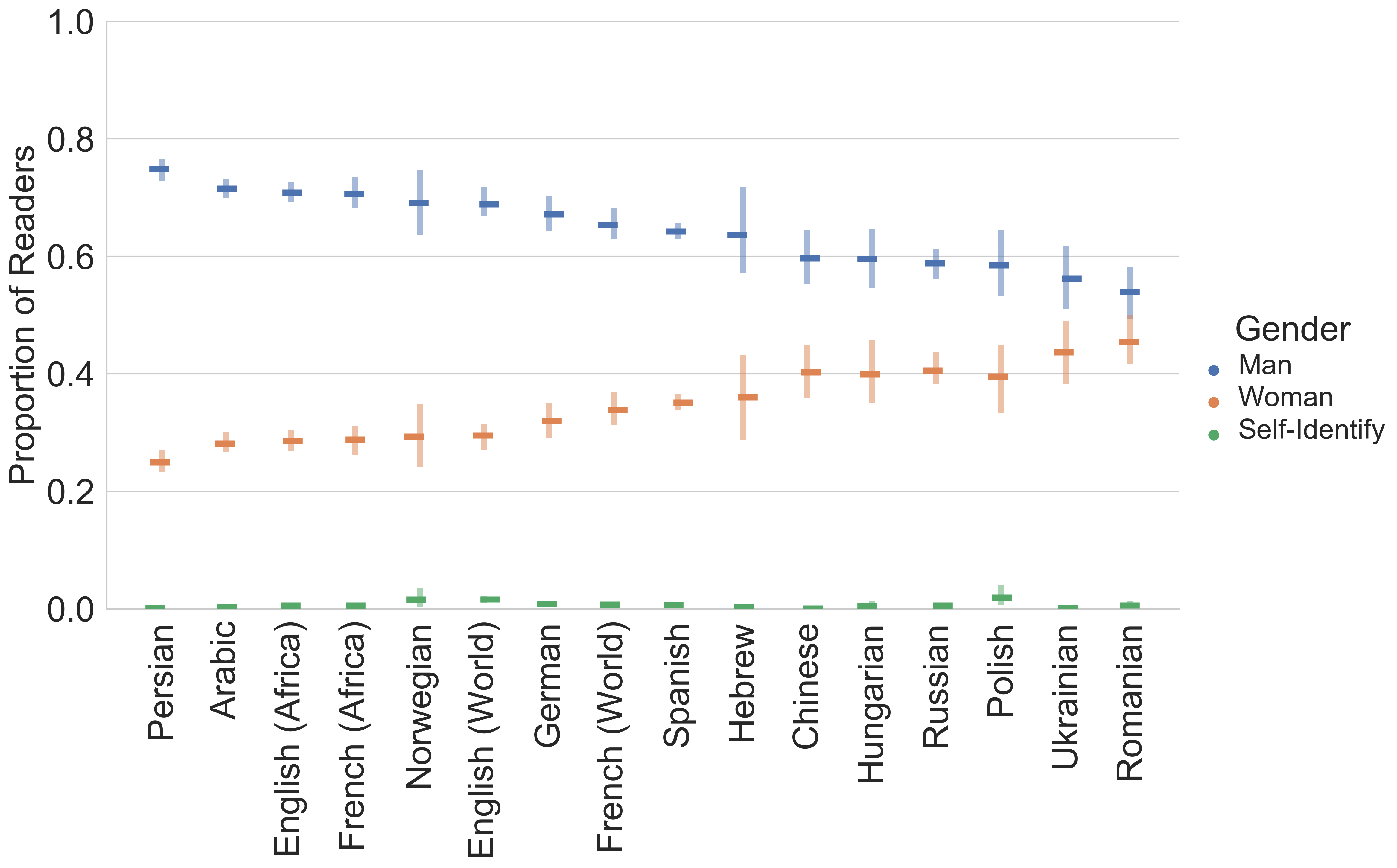}
        \caption{Gender of readers by language.}
        \label{fig:reader_gender_a}
    \end{subfigure}%
    ~ 
    \begin{subfigure}[t]{0.5\textwidth}
        \centering
        \includegraphics[width=\textwidth]{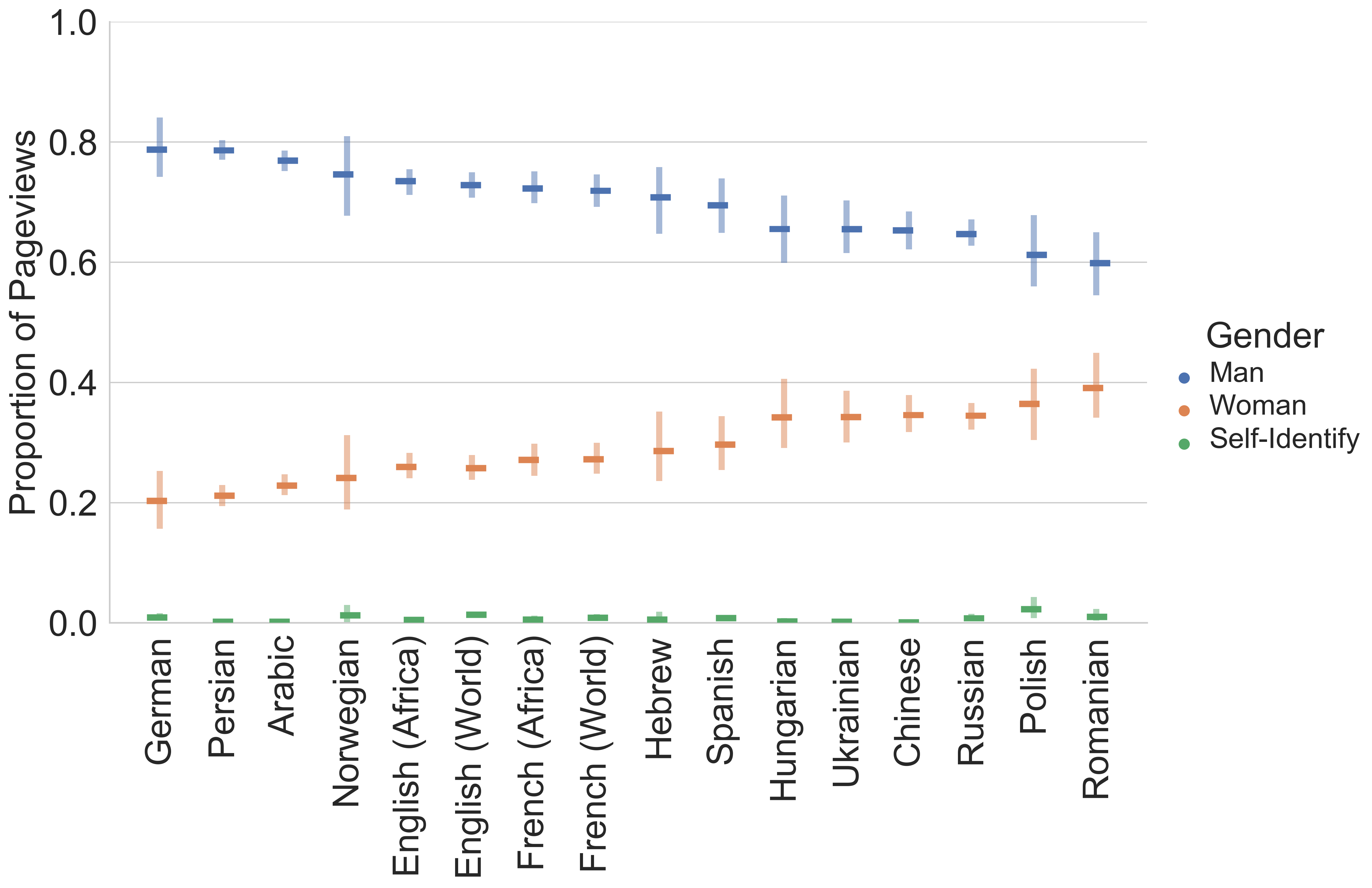}
        \caption{Gender of pageviews by language.}
        \label{fig:reader_gender_b}
    \end{subfigure}
    \label{fig:reader_gender}
    \caption{\textbf{Proportion of Wikipedia readers and pageviews by language and gender.} \textbf{(a)}, Proportion of readers who identified as women, men, or otherwise self-identified. The gap in gender representation is the highest in Persian Wikipedia (75\% men) and lowest in Romanian Wikipedia (54\% men). No language edition had a higher proportion of women readers than men. 
    \textbf{(b)}, Proportion of pageviews by language from readers who identify as men, women, or self-identify otherwise. The gap in gender representation in pageviews is the highest in German Wikipedia (79\% men) and lowest in Romanian Wikipedia (60\% men). However, even in the case of Romanian Wikipedia, this means that men generate 50\% more pageviews than women.}
\end{figure*}

We estimate that globally across all surveyed language editions of Wikipedia, about two-thirds of readers on any given day are men. While Figure~\ref{fig:reader_gender_a} shows substantial gender gap variation between languages (ranging from Romanian Wikipedia at 54\% men to Persian Wikipedia at 75\% men), it is notable that none of the language editions that we surveyed had a majority of women readers. This observation held true even for Norwegian Wikipedia readers\footnote{93\% of whom are in Norway; for more details, refer to Table~\ref{tab:survey_data} and Figure~\ref{fig:reader_country_gender} in the Appendix} for whom, informed by Norway's ranking as second in the list of countries with the smallest Global Gender Gap \cite{WEF:2020}, we had hypothesized gender parity in readership.

We also found that within individual language editions whose readership is distributed more evenly across multiple countries, different countries can have very different proportions of women readers. For instance, women comprised 36\% of readers in English Wikipedia in the United States, while they constituted only 24\% of readers in India (for more details refer to Figure~\ref{fig:reader_country_gender} in the Appendix).

Although men and women respondents shared many of the same demographic attributes (see Tables~\ref{tab:data_gender_x_education}, \ref{tab:data_gender_x_locale}, \ref{tab:data_gender_x_language} in the Appendix for education, locale, native language, respectively), we nevertheless found evidence of greater gender parity amongst younger readers. In 6 of the 16 surveys, women were significantly more likely than men to report being of age 18-24, and in no survey did men report to be in that age group significantly more often than women. For other age groups, we observed the opposite: Men were significantly more likely to report being older: 25-29 (4 of 16 surveys), 30-39 (7 of 16 surveys), 40-49 (4 of 16 surveys), and 50-59 (4 of 16 surveys), and in no survey did women report to be in one of those age groups significantly more often than men.

\subsection*{Women view fewer pages per reading session than men do}
By analyzing the pageview requests of survey respondents as available in Wikipedia's web server logs, we found that, on average, men had longer reading sessions than women; i.e. men read more articles on average when visiting Wikipedia than women. 
This observation holds for all 16 survey populations, but the magnitude of the difference varies (see Table~\ref{tab:data_session_pageviews_gender_language} in the Appendix). The largest difference was observed for the German Wikipedia edition, in which the mean session length of men ($3.94$ articles viewed) was almost twice the mean session length of women ($2.13$ articles viewed). By contrast, in Polish Wikipedia, the difference was only 13\% ($2.294$ articles viewed per session for men as compared to $2.021$ articles viewed per session for women).

As a consequence of these longer sessions, we observe an even stronger gender gap when considering visits to individual articles (pageviews) as opposed to visits by distinct individuals (readers). We estimate that overall, men comprise 67\% of readers but generate around 72\% of pageviews in the surveyed Wikipedia editions on any given day. Figure~\ref{fig:reader_gender_b} shows that results vary from German Wikipedia at 79\% of pageviews by men to Romanian Wikipedia at 60\%.

This overrepresentation of pageviews from readers who are men manifests strongly in the top-50 most-viewed articles during the week of the survey: we did not observe a single article that was viewed more often by women than men (see Table~\ref{tab:most_popular_articles} in the Appendix). The most-viewed article across all the survey respondents was for the \textit{Chernobyl disaster}\footnote{In English Wikipedia: \url{https://en.wikipedia.org/wiki/Chernobyl_disaster}. Received 2.5 million pageviews from all readers across all languages in the week of the survey: \url{https://pageviews.toolforge.org/langviews/?project=en.wikipedia.org&platform=all-access&agent=user&start=2019-06-26&end=2019-07-03&page=Chernobyl_disaster&sort=views&direction=1&view=list&page=Chernobyl\%20disaster}}, with 305 pageviews from survey respondents, of which 68\% came from men. The proportion of pageviews from men could range much higher, though, as with the \textit{2019 Africa Cup of Nations}\footnote{In English Wikipedia: \url{https://en.wikipedia.org/wiki/2019_Africa_Cup_of_Nations}}, with 243 pageviews from survey respondents, of which 85\% came from men. The lowest proportion of pageviews from men in these highly-viewed articles was to the article about the actor \textit{Mehdi Hashemi}\footnote{In English Wikipedia (most views came from Persian Wikipedia): \url{https://en.wikipedia.org/wiki/Mehdi_Hashemi_(actor)}}, where only 51\% of the 155 pageviews from survey respondents were from men.

\subsection*{Men and women visit Wikipedia for similar reasons}
In the survey, we also asked Wikipedia readers across language editions about their information need (i.e. the depth of information they sought when on Wikipedia: a specific fact or a quick answer, an overview, or an in-depth read) as well as their motivation for visiting the site. We then analyzed whether there are differences between men and women with respect to information need and motivation.

\begin{figure}[ht]
\centering
\includegraphics[width=1\textwidth]{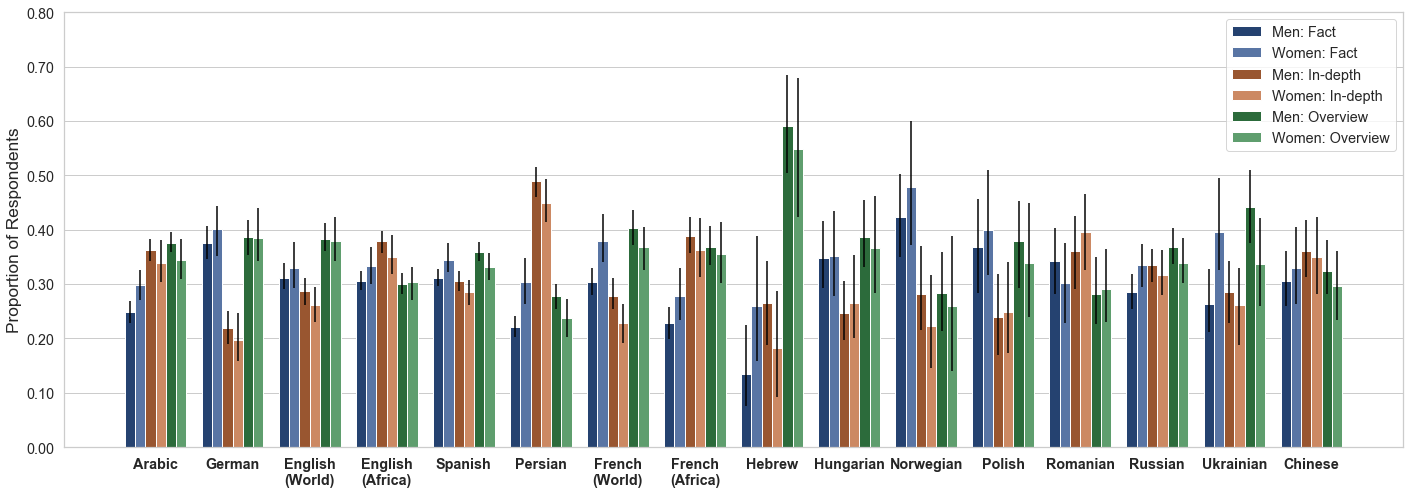}
\caption{\textbf{Information needs per gender across Wikipedia languages.} We report the proportion of men and women for each category of information need: (i) facts, (ii) overview and (iii) in-depth information. The data shows that men and women have very similar information needs, even in languages like Persian, Hebrew, and Norwegian, which have much different results than the other surveyed languages. Though women are more likely to be looking for a a fact in 7 of the 16 languages, a reader's gender is not a strong determinant of the depth of information sought. 99\% confidence intervals are shown.}
\label{fig:gender_infoneed}
\end{figure}

At a high level, we found that men and women reported similar information needs (see Figure~\ref{fig:gender_infoneed}). For 9 of the 16 surveys, there were no significant differences between men and women for any reported information need. For the other 7, women reported looking for a fact significantly more than men. Differences between languages, however, were often much larger than differences between men and women; \emph{e.g.}, in Arabic Wikipedia, women reported looking for facts 30\% of the time, while men reported looking for facts 25\% of the time; but in German Wikipedia, men and women both looked for a fact about 10 percentage points more often: 40\% of women and 38\% of men.

Considering reader motivations to read Wikipedia (see Table~\ref{tab:data_gender_x_motivation} in Appendix), the same high-level trends were seen amongst men and women (again with a few exceptions). We saw little evidence of consistent gender differences in any of the following motivations: conversation (2/16 surveys with significant differences), media (3/16), intrinsic learning (5/16), or help making a personal decision (2/16). In 12 of the 16 survey populations, however, men were more likely to report boredom or randomly reading Wikipedia for fun as their motivation for browsing Wikipedia. Men also were more likely to be motivated by a current event for checking Wikipedia in 7 of the 16 survey populations. Finally, women, reported work or school as a motivation significantly more often than men in 7 of the 16 survey populations. There were no significant differences in the other surveys for boredom, current events, or work and school.

\subsection*{Men and women exhibit specific topical preferences}

\begin{figure}[ht]
\centering
\includegraphics[width=1\textwidth]{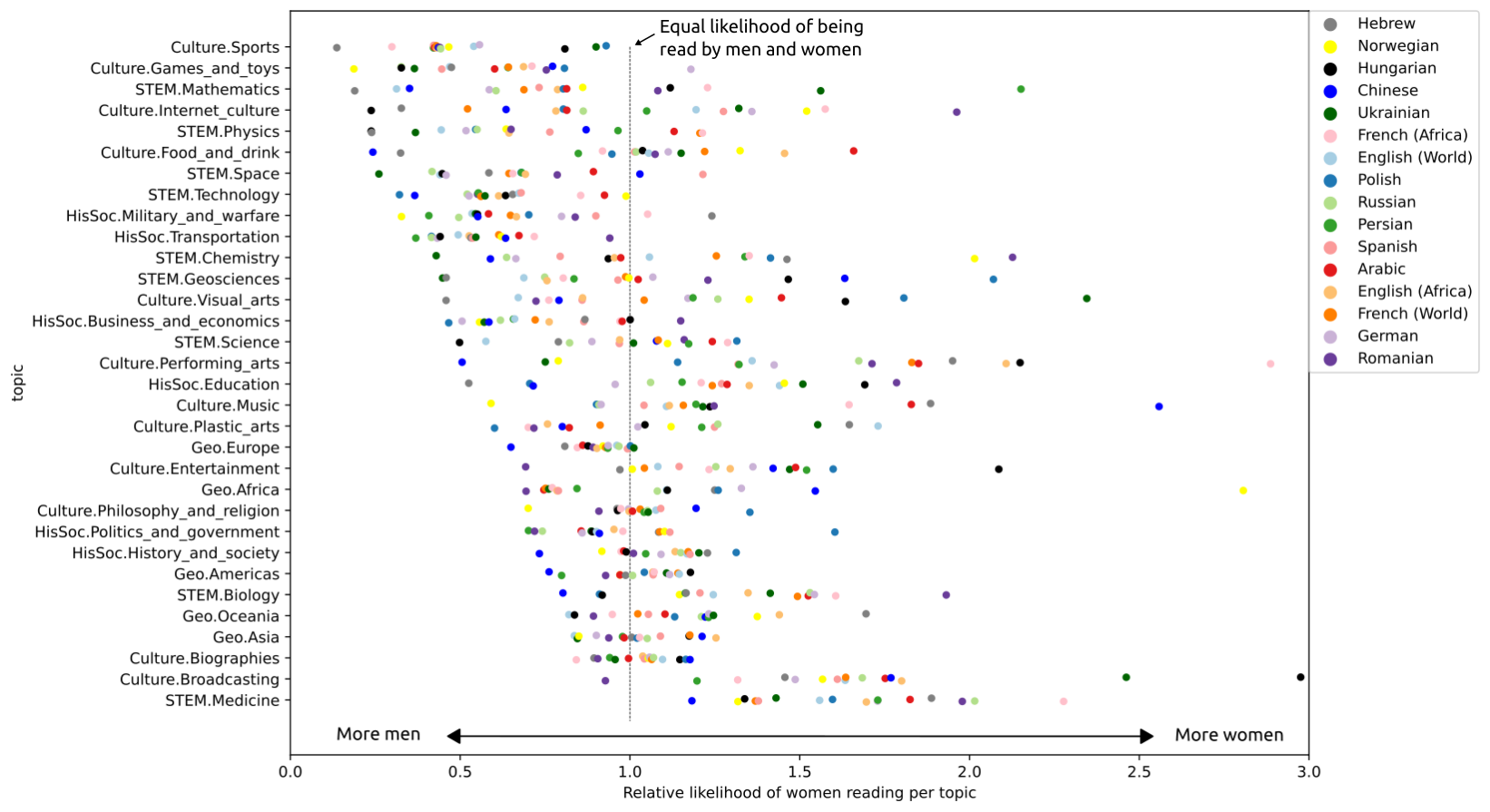}
\caption{\textbf{Topical interest in Wikipedia by gender and language.} We report the relative frequency of reading about various topics on Wikipedia for women and men per survey---i.e. $P(\text{topic}|\text{woman}) / P(\text{topic}|\text{man})$. In particular, if men and women are equally likely to view a topic, this is represented as 1 on the x-axis while a value of 1.5 would indicate that women are 50\% more likely to read about that topic than men and a value of 0.66 would indicate that men are 50\% more likely to read about the topic than women. For all topics, women still generate fewer pageviews than men given the overall imbalance of pageviews from men. We see that in topics like Sports and Games, the share of women interest (on average) is low for all languages surveyed, with exception of women reading about \textit{Sports} in the German Wikipedia. Topics like \textit{Medicine} receive relatively more interest from women. In the majority topics, however, like \textit{Philosophy and Religion} or most geographic topics, the attention between men and women either has no significant differences in any survey or just a few significant differences but mixed in their direction. We truncate the x-axis at 3, so two language-topic pairs do not appear: Norwegian for \textit{STEM.Space} has a value of $4.9$ and Polish for \textit{Culture.Broadcasting} has a value of $4.3$.}
\label{fig:gender_topic_womenshare}
\end{figure}

While many topics on Wikipedia were read equally frequently by men and women, we found that across the 16 surveys, there were both topics that were read more consistently by men (\emph{e.g.}, sports or technology) and topics that were read more consistently by women (e.g., broadcasting or medicine).
In Figure~\ref{fig:gender_topic_womenshare} we visualize the gender skew for each topic in each language edition.

However, it is important to note that even for topics such as broadcasting (\emph{e.g.}, television shows) or medicine, which women were more likely to read than men, men still generated the majority of pageviews. For instance, articles about the Chernobyl miniseries\footnote{In English Wikipedia: \url{https://en.wikipedia.org/wiki/Chernobyl_(miniseries)}} were the 13\textsuperscript{th}-most-read articles by the survey population (122 times across all of the languages). Although the articles are categorized as broadcasting and women were more likely to read about the Chernobyl miniseries than men, 68\% of the pageviews still came from men. This can be explained by our earlier observation that men have a higher general frequency of reading Wikipedia and longer reading sessions.

\begin{figure}[ht]
\centering
\includegraphics[width=1\textwidth]{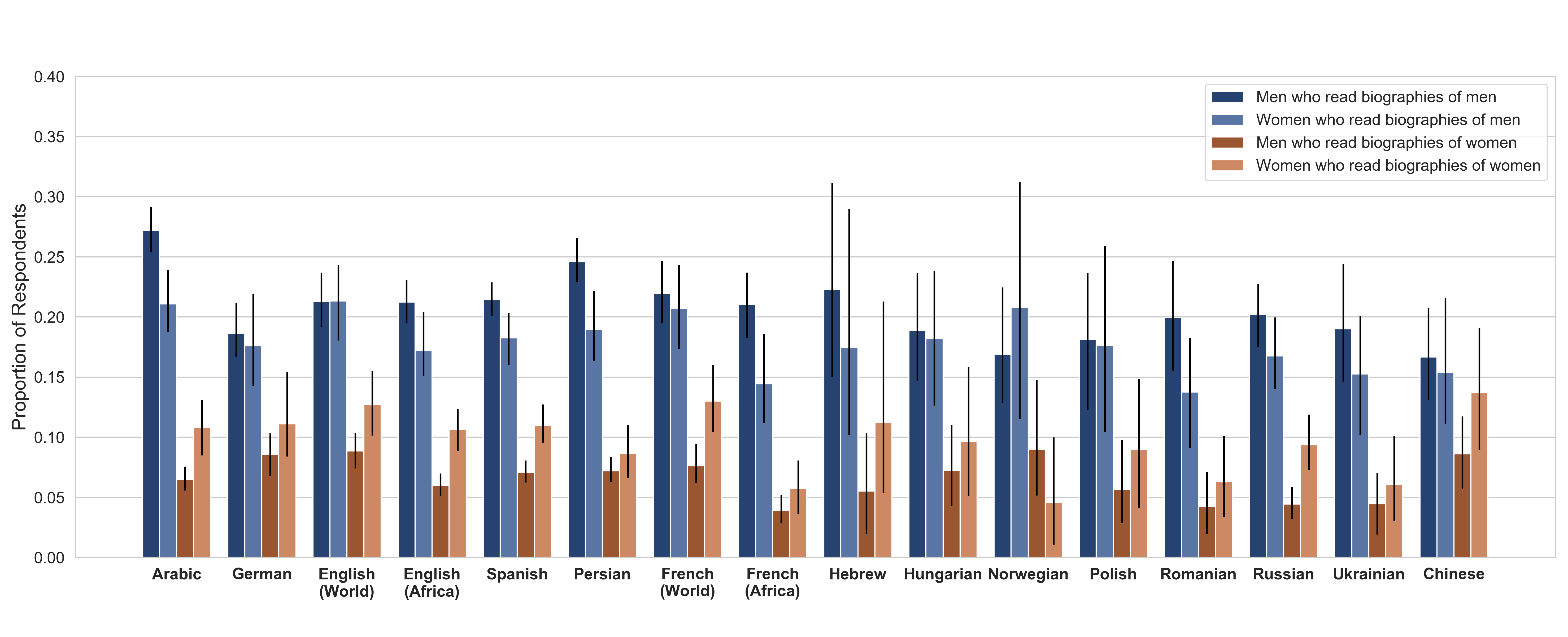}
\caption{\textbf{Interest in Wikipedia biographies about men and women by reader gender.} We report the proportion of men and women Wikipedia readers who read biographies about men or biographies of women. Wikipedia has many times more biographies of men (approximately 8 times more depending on the language). Therefore, it is perhaps surprising that biographies of women are read at nearly half the rate as biographies of men. Approximately 35\% of reader sessions include an article that is a biography. 99\% confidence intervals provided.}
\label{fig:gender_biography}
\end{figure}

An interesting observation with respect to topical interest is that there was substantial self-focus in the most popular topic on Wikipedia: biographies (about 35\% of pageviews). Figure~\ref{fig:gender_topic_womenshare} reveals that biographies (Culture.Biography) are balanced in how often they were viewed by men and women overall.
A different picture emerges when considering the gender of the person who is described in the viewed biography. Figure~\ref{fig:gender_biography} demonstrates clear self-focus bias: men were more likely to read biographies of men than women were (7 of 16 survey populations), whereas women were more likely to read biographies of women than men were (7 of 16 survey populations, with the other surveys showing no significant differences in both cases).

\section*{Discussion}
The surveys in this paper provide comprehensive data from a wide variety of languages that demonstrate a consistent gender gap amongst Wikipedia readers and evidence of behavioral differences between men and women for session length and topical interests. These results indicate that there are still large gaps in usage of Wikipedia and likely barriers to equal access to encyclopedic knowledge across gender identities. Investigating potential causes for this gender gap in readership will be a challenging and crucial task for future research.

Although the focus of our surveys was on measuring the readership gender gap as opposed to investigating its underlying causes, our comparison of reading behavior between men and women suggests possible mechanisms. In particular, our research raises questions about the impact of content gaps on who reads Wikipedia. The gender gap in Wikipedia content is well-documented~\cite{graells2015first,lam2011wp,klein2016monitoring,reagle2011gender,wagner2015s}; \emph{e.g.}, less than 20\% of biographies in English Wikipedia are about women~\cite{wade2018we}. Prior research has suggested that women receive less value from Wikipedia~\cite{lim2010gender,garrison2015getting}, likely in part due to these content gaps. We did not see any indications in our data that women tended to read lower-quality content on Wikipedia. In particular, there were no consistent differences in the average length of articles, number of headings, or other features that are related to article quality~\cite{warncke2013tell}. 
We observed, however, that women showed more interest than men in topics such as biographies of women, medicine, and broadcasting. Missing, low-quality, or otherwise biased content in these areas would therefore have an outsized impact on women readers. This finding highlights the importance of work conducted by Wikipedia volunteers to increase content about women in Wikipedia~\cite{wade2018we,halfaker2017interpolating} and calls for further work in this space.

Wikipedia readership gaps relate directly to the gender gaps among Wikipedia contributors. The pipeline of participation suggests that the gender gap observed in contributors is in fact a function of gaps that appear in awareness of Wikipedia, readership of Wikipedia, awareness of the ability to edit Wikipedia, and only then retention of Wikipedia editors.~\cite{shaw2018pipeline} Our surveys provide global data about the state of gender gaps in readership. If these gaps are not addressed, there will continue to be large gender gaps amongst Wikipedia contributors~\cite{hill2013wikipedia,ford2017anyone,antin2011gender,lam2011wp,collier2012conflict,sichler2014gender}, which will likely reinforce associated content gaps due to self-focus bias~\cite{hecht2010tower,das2019gendered}.

An important distinction should be made about the readership measured in these surveys: while the results of this study show that the readers of many Wikipedia language editions on any given day are predominantly men, this does not necessarily imply that fewer women read Wikipedia. In fact, a number of surveys have found that women are just as likely as men in many regions to identify as readers of Wikipedia, though less likely to have read Wikipedia in the recent past\cite{glott2010wikipedia,pew2011wikipedia,protonotarios2016similar}; \emph{e.g.}, ``do you read Wikipedia?'' as compared to ``did you read Wikipedia yesterday?''. Together, these surveys and past surveys indicate that, while women may be equally aware of Wikipedia, on average they visit less frequently and, as we have shown, read fewer pages when they do visit. The reader behavior analyses we presented here further offer new insights into how these gaps might arise and manifest in research and tools that are built with reader data. This data hopefully can stand as a baseline and motivation for increased efforts to address these gaps.

Recommender systems, tools that provide editors with ranked lists of articles to create or edit~\cite{laxstrom2015content,wulczyn2016growing,cosley2007suggestbot}, constitute an important approach to closing content gaps. These tools often use pageviews as a metric for ranking the recommendations. In this study we found that men generate a large majority of pageviews on Wikipedia and an even larger proportion of pageviews for many popular articles related to topics such as sports. Researchers and developers of these recommender systems should take precautions to avoid prioritizing content that is predominantly read by men, as opposed to content that is inclusive of all genders. The gender gap amongst readers, however, is smaller than the gender gap that has been observed amongst editors with estimates that only 9\% of Wikimedia project contributors identify as women.\footnote{\url{https://meta.wikimedia.org/wiki/Community_Insights/2018_Report\#Diversity_of_contributors_on_the_Wikimedia_projects_seems_to_remain_unchanged.}} 
This suggests that the interests of readers can still be a valuable signal for guiding the prioritization of content to be more representative of what people (and not just existing readers) need or want to know.

Lastly, Closing content gaps is not a panacea as evidenced by prior research on Welsh Wikipedia, where a majority of the biographies are about women~\cite{lubbock_welsh_2016}, a majority of Welsh speakers are women,\footnote{\url{https://statswales.gov.wales/Catalogue/Welsh-Language/Annual-Population-Survey-Welsh-Language/welsh-skills-by-age-sex}} but readership is still heavily skewed towards men~\cite{welsh2017}. We refer interested readers to a list of other possible causes for the gender gap in readership\cite{morgan2019}. Our findings lay the foundation for identifying pathways toward improved knowledge equity in the usage of online encyclopedic knowledge and we encourage the research community to build on the learnings from this study and help address Wikipedia's gender gap.

\section*{Methods}
We collected in-situ survey responses from 65,031 users reading Wikipedia, alongside server logs of their click sessions during which the survey was completed.\footnote{Per our survey privacy policy: \url{https://foundation.wikimedia.org/wiki/2019_Wikipedia_Demographics_Survey_Privacy_Statement}}
The survey was run worldwide in 14 Wikipedia language editions (see Table~\ref{tab:survey_data}) from 26 June 2019 until 3 July 2019.\footnote{With the exception of Polish Wikipedia, which was run from 26 September 2019 through 31 October 2019.} We selected the language editions with the following considerations in mind: diversity of language family, geographic diversity (as far as primary location of readers), and diversity of size of readership with the constraint that the language must receive sufficient pageviews to support the survey. In addition to our initial set of languages, we also included languages by requests of Wikipedia volunteers. For the globally spoken languages English and French, in addition to sampling users worldwide, we included a separate sampling procedure that specifically targeted Wikipedia readers in Africa (geolocated based on their IP addresses) to receive sufficient data to study potential regional differences in usage for these editions. This led to a grand total of 16 surveys across 14 languages that receive 78\% of the monthly pageviews across all language editions of Wikipedia.

\begin{table}[ht]
\small
\centering
\begin{tabular}{l|p{2.2cm} p{2.2cm}p{8cm}}
Survey & Response Count (over 18) & Response Count (under 18) & Countries with at least 500 responses \\
\hline
Arabic (ar) & 7741 & 1733 & Saudi Arabia~(2077), Egypt~(1245), Iraq~(726) \\
German (de) & 4144 & 540 & Germany~(3457) \\
English (en -- Worldwide) & 6181 & 1167 & United States~(1816), India~(1673) \\
English (en -- Africa) & 8043 & 1159 & South Africa~(2257), Nigeria~(1508), Kenya~(930), Egypt~(720) \\
Spanish (es) & 11897 & 4985 & Spain~(2125), Mexico~(1942), Argentina~(1834), Colombia~(1197), Peru~(891), Chile~(619) \\
Persian (fa) & 7036 & 1867 & Iran~(5667) \\
French (fr -- Worldwide) & 4401 & 1126 & France~(3145) \\
French (fr -- Africa) & 3122 & 799 & Morocco~(705), Algeria~(622) \\
Hebrew (he) & 586 & 353 & Israel~(556) \\
Hungarian (hu) & 1216 & 412 & Hungary~(1019) \\
Norwegian (no) & 737 & 180 & Norway~(689) \\
Polish (pl) & 688 & 284 & Poland~(636) \\
Romanian (ro) & 1336 & 453 & Romania~(1157) \\
Russian (ru) & 4565 & 1253 & Russia~(2835), Ukraine~(685) \\
Ukrainian (uk) & 1148 & 368 & Ukraine~(1027) \\
Chinese (zh) & 2190 & 814 & Taiwan~(1310) \\
\end{tabular}
\caption{\textbf{Survey response counts and country breakdowns.} For each survey, we provide the number of responses from individuals who indicated that they were over the age of 18 (what we analyze in this research) and individuals who indicated that they were under the age of 18 (so they did not complete the survey). Additionally, for each survey, we provide the countries with at least 500 responses and how many respondents they had who were over the age of 18. For French and English, we cannot actually distinguish between how many readers under the age of 18 came from Africa as opposed to the other continents. As such, we divide the under-18 respondents in proportion to how many over-18 respondents there were for each survey.}
\label{tab:survey_data}
\end{table}

\subsection*{Survey Questions}
The survey solicited information about the respondents' demographics (age, gender, education, locale, native language) and their reasons for reading Wikipedia (motivation, information need, prior knowledge).
Via server logs, the survey responses were enriched with information about the situational context (\emph{e.g.}, geography, time of day) and the user's behavior while reading Wikipedia (\emph{e.g.}, session length, topics read, whether readers switched language editions while reading).

The survey questions were designed with the goal of balancing simplicity, privacy, and applicability to a global audience. Questions were adapted from prior, validated surveys where possible. In particular, we reused questions about motivation and information need from existing publications that targeted these topics specifically~\cite{singer2017we,lemmerich2019world}, while for demographic questions we adapted questions from multiple sources including the International Social Survey Program~\cite{edlund2019issp} and previous surveys on Wikipedia.

Utilizing the taxonomy of Wikipedia readership use-cases~\cite{singer2017we}, we asked respondents three multiple-choice questions, based on a validated taxonomy of Wikipedia reading behavior:
\begin{enumerate}
    \item \textit{I am reading this article because}
(a)~I have a work or school-related assignment;
(b)~I need to make a personal decision based  on  this  topic  (\emph{e.g.},  buy  a  book,  choose  a  travel destination);
(c)~I want to know more about a current event (\emph{e.g.},  a  soccer  game,  a  recent  earthquake,  somebody's death);
(d)~the topic was referenced in a piece of media (\emph{e.g.}, TV,  radio,  article,  film,  book); 
(e)~the  topic  came  up  in  a conversation;
(f)~I am bored or randomly exploring Wikipedia for fun;
(g)~this topic is important to me and I want to learn more about it (\emph{e.g.}, to learn about a culture). Users could select multiple answers for this question.
    \item \textit{I am reading this article to}
(a)~look up a specific fact or to get a quick answer;
(b)~get an overview of the topic;
(c)~get an in-depth understanding of the topic.
    \item \textit{Prior  to  visiting  this  article I was}
(a)~already  familiar with the topic;
(b)~not familiar with the topic, and I am learning about it for the first time.
\end{enumerate}

Free-form answers were also allowed, but the vast majority of respondents chose from the pre-defined answers, suggesting that the taxonomy developed in the earlier study~\cite{singer2017we} remains comprehensive.

The five demographic questions were based on past surveys of readers and factors known to affect readership: gender, age, education level, locale (urban vs.\ rural), and native language. For this paper, we focused on gender. We applied best practices of allowing respondents to select from many identities or self-identify via inclusive language\footnote{\url{https://www.morgan-klaus.com/gender-guidelines.html}} within the constraints of the Google Forms platform. We provided pre-set answers of ``Man'', ``Woman'', and ``Prefer not to say'' along with a free-text ``Other'' option. Although a number of respondents identified their gender as non-binary, the usage of ``Other'' was too low to perform statistical analyses (417 respondents across all the surveys).

The five demographic questions were as follows:
\begin{enumerate}
    \item \textit{What is your age?}
(a)~18-24 years;
(b)~25-29 years;
(c)~30-39 years;
(e)~40-49 years;
(f)~50-59 years;
(g)~60 years and older;
(h)~Prefer not to say.
    \item \textit{What is your gender?}
(a)~Woman;
(b)~Man;
(c)~Prefer not to say;
(d)~Other... <open-text>.
    \item \textit{How many years (full-time equivalent) have you been in formal education? Include all primary and secondary schooling, university and other post-secondary education, and full-time vocational training, but do not include repeated years. If you are currently in education, count the number of years you have completed so far.}
(a)~I have no formal schooling;
(b)~1-6 years;
(c)~7 years;
(d)~8 years;
(e)~9 years;
(f)~10 years;
(g)~11 years;
(h)~12 years;
(i)~13 years;
(j)~14 years;
(k)~15 years;
(l)~16 years;
(m)~17 years;
(n)~18 years;
(o)~>18 years;
(p)~Prefer not to say.
    \item \textit{Would you describe the place where you live as...}
(a)~A farm or home in the country;
(b)~A country village;
(c)~A small city or town;
(d)~The suburbs or outskirts of a big city;
(e)~A big city;
(f)~Prefer not to say.
    \item \textit{What is your native language?}
<list of Wikipedia languages in their native script with language codes with the survey language at the top of the list>
    \item \textit{What is your second native language?}
(a)~I do not have a second native language;
(b)~Other... <open-text>.
\end{enumerate}

The survey was designed and piloted in English and then translated into the 13 other languages with the assistance of native speakers, comprised primarily of Wikipedia volunteers in these languages. Respondents could skip any of the demographic questions, though in practice response rates for those questions were greater than 90\% and in most cases >95\% of survey participants for all surveys. A link to the survey was displayed within Wikipedia articles as the reader was browsing (details below under ``Survey Sampling''). This means that the three motivation and information need questions were answered in the context of a particular article the respondent was reading. For legal reasons, we additionally provided an initial screening question that removed individuals under the age of 18 (see Table~\ref{tab:survey_data} for details on how many respondents were screened out per language).

\subsection*{Survey Sampling}
\begin{figure}[ht]
\centering
\includegraphics[width=0.9\textwidth]{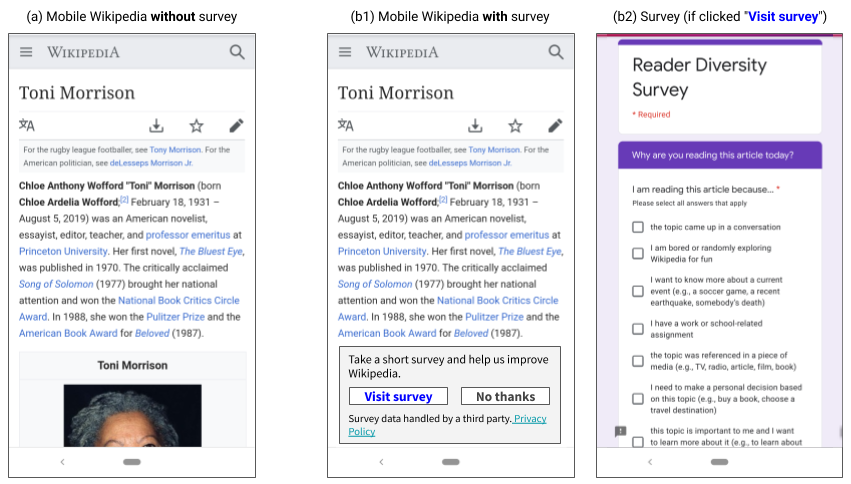}
\caption{\textbf{Participant recruitment for the survey on Wikipedia.} Illustration of how survey invitations appeared on a mobile Wikipedia article. Panel (a) shows how an article is normally displayed on mobile Wikipedia. Panel (b1) shows where the survey invitation would be inserted within the article. The survey invitation behaves like article content so the reader can easily scroll past the survey. Panel (b2) shows one of the pages from the survey if the reader clicked ``Visit survey'' (the initial page asks the reader if they are above the age of 18). Display for desktop Wikipedia not shown but similar.
}
\label{fig:quicksurveys}
\end{figure}

A prompt that asked users to participate in a survey in order to help us improve Wikipedia was shown using the MediaWiki QuickSurveys extension\footnote{\url{https://www.mediawiki.org/wiki/Extension:QuickSurveys}} as a box embedded toward the top of articles on the desktop and mobile versions of Wikipedia (see Figure~\ref{fig:quicksurveys}).
Users who chose to participate were sent to a language-specific questionnaire hosted on Google Forms.

Sampling rates were set on a per-language basis and based on predictions on the number of daily pageviews expected in each language (ranging from 1:2 for Norwegian Wikipedia to 1:98 for English Wikipedia). For English and French Wikipedia, we applied two separate sampling procedures: one general random sample of global readership (referred to as ``Worldwide'', this is how we sampled the other languages, too), and another one with filters that specifically sampled readers from African countries (referred to as ``Africa''). The first time a browser visited a Wikipedia language edition with an active survey, a random hash ID was generated that deterministically indicated whether the browser would see the survey or not. The hash ID was stored in the browser and remained there unless cookies were refreshed. If the survey was taken or dismissed, the hash was adjusted to indicate this and the survey was no longer shown on that browser. Thus, anyone using that particular browser would continue to see the survey on each Wikipedia article they viewed in that language until they either took the survey, dismissed it, or cleared their cookies. The survey was not shown on browsers with Do-Not-Track enabled. This sampling strategy was simple and preserved privacy (as all logic occurred on the client side), but had the limitation that an individual could potentially see and respond to the survey on multiple browsers (though in practice, we saw no evidence of this).

We also acknowledge that the identification of individual reader sessions cannot be guaranteed to always be 100\% accurate---\emph{e.g.}, there might be multiple individuals sharing one computer or a device's IP address could change during a single session. Neither scenario would be detected by our approach, but, even if these situations occurred, there is no reason to believe this would introduce a strong systematic bias towards any particular gender identity in the results.

\subsection*{Survey Linking}
When an individual responded to the survey, a unique code was passed through the survey that we use to directly link that individual's response to the Wikipedia article that they were reading when they took the survey. We could then reconstruct that reader's broader session of pageviews associated with the survey response, under the assumptions that the individual who took the survey was the only reader of Wikipedia on the browser in which the survey was taken in that period of time, that all of that reader's pageviews came from a single device and browser, and that that device was associated with a single IP address. We define a reader session as consecutive pageviews with no more than one hour between subsequent pageviews based on past research~\cite{halfaker2015user}. We limit our analyses to just the reader session in which the survey was taken to reduce the risk of violating our assumptions---\emph{e.g.}, over longer time periods, we would expect mobile devices to change IP addresses.

\subsection*{Bias and Validity}
Following the procedure from previous studies of Wikipedia readership through surveys~\cite{singer2017we,lemmerich2019world}, we adjusted for potential biases in the survey responses to the best of our ability given the availability of features that may capture biases. To this end we used \emph{inverse propensity weighting}~\cite{austin2011introduction,lunceford2004stratification} utilizing behavioral data extracted from the pageview logs.
We compared the reading behavior of the survey respondents to the reading behavior of a representative sample of readers from that language edition. We learned weights that rebalance the survey-respondent population for a given language so that it has similar observed characteristics to the full reader population for that language edition.
In particular, we controlled for the following covariates:
\begin{itemize}
    \item \textbf{contextual features}: day of week, time of day, country, continent
    \item \textbf{reader session features}: session time length, session access method, average time between pageviews, session referer class, where in session survey was taken, number of Wikipedia languages viewed, number of pageviews, reader logged-in, reader viewed Wikipedia main page\footnote{\emph{e.g.}, English: \url{https://en.wikipedia.org/wiki/Main_Page}}
    \item \textbf{article demand features}: average and entropy of pageviews to articles read, average and entropy of number of languages each article is available in, pageviews to article where survey was taken
    \item \textbf{article topic features}: biography of a man, biography of a woman, article has latitude-longitude coordinates, article is for an event with a point-in-time, article topics
    \item \textbf{article quality features}: average and entropy of length of articles read, average infonoise \cite{warncke2013tell}, average number of second-level and third-level headings, average number of templates in article, average number of references, average and entropy of number of internal links, average number of external links
\end{itemize}

With inverse propensity score weighting, each survey response gets assigned a weight that is the inverse of the predicted likelihood that that person would respond to the survey based on their reading behavior. For this prediction, we used a gradient boosting classifier as implemented in the scikit-learn Python library, using default parameters.\footnote{\url{https://scikit-learn.org/0.19/modules/generated/sklearn.ensemble.GradientBoostingClassifier.html}}
To reduce the effect of strong outliers, the top 5\% of weights were trimmed to the 95\% threshold~\cite{potter1993effect}, see also~\cite{lee2011weight}. The application of trimming did not change any of the main trends we discuss.

By analyzing the gradient boosting classifier and the resulting weights, we observed that the most noticeable effect of weighting was with respect to the number of articles viewed by a reader: readers who view more pages are also more likely to respond to the survey and thus are weighted lower in the results than readers who view a single page. In practice, this correction only shifted the results by at most 6\%, with, \emph{e.g.}, consistent shifts for our estimates of readers who are motivated by boredom (overrepresented in the raw results) and who identify as women (underrepresented in the raw results). These debiased results are used for all analyses with the exception of simple count-based results---\emph{e.g.}, number of responses per country for a survey or total pageviews from survey respondents to a given article.

A number of additional robustness checks were implemented that validated the survey results. The three motivation-related questions in the survey were also asked of Wikipedia readers in several of the 14 languages in 2017 using very similar methods~\cite{lemmerich2019world}. We saw for those questions that the results had largely remained stable between 2017 and 2019. For Russian and English, we also ran the survey for one week in June 2019 (reported here) and again for one month in September 2019 (not reported). Even with the longer sampling time-frame, which would give infrequent readers more opportunities to see and respond to the survey, we saw nearly identical results (with small differences that were largely explained by seasonality).

\subsection*{Article Topics}
For our analyses, we compared the types of content that men and women read on Wikipedia. We faced the challenge that our surveys were deployed in 14 languages and the respondents of these surveys viewed articles in a total of over 100 languages over the course of their reading sessions. Traditional topic modeling would be computationally intensive and require extensive hand-labeling of topics in many languages to provide complete coverage of the dataset. To assign consistent topics across all of these languages, we instead relied on Wikidata, which contains language-independent structured data about the concepts covered in Wikipedia articles. Specifically, we employed two methods of categorizing the articles that the survey respondents viewed in our dataset: (1)~we deterministically identified articles as biographies based on their associated Wikidata item, which specifies whether the article is about a person and, if so, what that person's gender identity is, and
(2)~we trained a model to predict which of 44 high-level topics (such as \textit{STEM.Medicine} or \textit{Geography.Europe}) apply to any given Wikipedia article using a topic classification model trained on Wikidata attributes. The topics for this classification task were extracted from earlier research~\cite{asthana2018few} on building a taxonomy for Wikipedia article topics that spreads across four high-level categories (STEM, Geography, History and Society, Culture), each of which have many sub-categories. Across these topics, we achieved a micro F1 score of 0.811 and macro F1 score of 0.643.\footnote{The model can be explored at \url{https://wiki-topic.toolforge.org/\#wikidata-model}}

\subsection*{Statistical Analyses}
We present the results of simple comparison-based analyses---\emph{e.g.}, whether the proportion of women aged 18-24 is significantly different than the proportion of men aged 18-24---in the body of this work as they most directly reflect the composition of Wikipedia readers. We additionally provide a more comprehensive regression-based analysis that controls for the other main covariates in the Appendix (see Table~\ref{tab:regression}). Specifically, for each survey, we run a logistic regression\footnote{Using the \textit{glm} package in R with our debiased sample weights} for that survey's responses that has as the dependent variable a binary variable that indicates whether the respondent identified as a man and as the independent variables the categorical variables for topic, information need, country, age, education, locale, and native language.
These regressions provide deeper insight into whether the correlations we see in our simple analysis can be explained by other observed factors. While in all cases we see support for our comparison-based conclusions, a lack of significance after controlling for other variables would in no way detract from the conclusion that women are significantly underrepresented in the global readership of Wikipedia; this would merely indicate that the reasons why we see lower readership among women are likely as complex as the causes hypothesized by research on the gender gap amongst Wikipedia editors~\cite{lam2011wp,protonotarios2016similar,hinnosaar2019gender}.

For all statistical analyses, we rely on 99\% confidence intervals for the determination of what results we consider to be significant. For the comparison-based analyses, we computed 99\% confidence intervals through bootstrap resampling of the weighted survey responses with 400 iterations. For the regression-based analyses, we use standard $p$-values but only report results as significant at $p<0.01$. Given that we have sixteen surveys and many variables under study, we face the challenge of multiple comparisons. We do not make Bonferroni corrections or other such adjustments but instead rely on the 99\% confidence level and presence of 16 independent surveys to act as a natural check on false positives---namely, we do not report trends as significant unless we both see them significant across at least 5 surveys along with no significant trends in the opposite direction from any of the surveys.

With the exception of our estimates of the proportion of women in worldwide readership and pageviews, we avoid directly combining our results---\emph{e.g.}, a multi-level regression model that might produce a single significance value for a given covariate. For the former, it is clear that the survey results should be weighted by the proportion of pageviews each language produces as we seek a single number representative of the current state of global readership. As such the results for English Wikipedia (50\% of pageviews) heavily influence the results while minor language editions such as the Romanian Wikipedia (0.2\% of pageviews) only have little weight. For more advanced analyses, giving proportionally more weight to the English Wikipedia than to other language editions simply because it has more readers would obfuscate the findings for smaller language editions. Since the focus of this work was to investigate gender differences \emph{worldwide}, we run our analyses separately for each survey and give each uniform weight in discussing the global trends that we observe.

\section*{Acknowledgements}
We thank Bahodir Mansurov, Wikimedia Foundation (WMF) Readers Web and Legal teams, and the following volunteers for helping to make these surveys happen: Abbad, Amire80, Yury Bulka, Adélaïde Calais WMFr, Kaganer, Liang-chih ShangKuan, Jon Harald Søby, Strainu, Tgr, Nettrom. We thank Clemens Lechner (GESIS - Leibniz Institute for the Social Sciences) for providing advice on the selection and harmonization of survey questions.

\section*{Author contributions statement}
F.L., M.S., R.W. and L.Z. defined the overall direction of the research.
I.J. took the lead in finalizing the survey questions and all authors contributed to the discussion. 
I.J. and L.Z. worked with the Wikipedia volunteer community to translate the survey questions.
I.J. conducted the surveys.
I.J., D.S., and F.L. analyzed the results.
All authors contributed to the manuscript and reviewed it. All authors approved the final version of the manuscript.

\section*{Additional information}

\textbf{Competing interests}
The authors declare the following competing interests: three of the authors are employed by the Wikimedia Foundation, the 501(c)3 non-profit organization that operates Wikipedia and its sister projects. Two of the authors are Formal Collaborators of the Research team at the Wikimedia Foundation. One of the authors is a Research Fellow at the Wikimedia Foundation. Formal Collaborations and Research Fellow positions are volunteer positions.

\section*{Appendix}
This appendix contains tables and figures that provide additional supporting evidence for claims made in the paper and the cross-tabulation of gender with the other survey questions not discussed in the body of the paper.

\begin{table}[ht]
\centering
\begin{tabular}{l|c c}
Survey & Avg.\ \# Requests (Men) & Avg.\ \# Requests (Women) \\
\hline
Arabic             & 2.465 [2.350-2.614]   & 1.862 [1.753-2.007] \\
German             & 3.935 [3.128-5.513]   & 2.127 [1.915-2.392] \\
English (World)    & 2.853 [2.706-3.046]   & 2.355 [2.166-2.598] \\
English (Africa)   & 2.424 [2.304-2.544]   & 2.122 [1.997-2.337] \\
Spanish            & 2.791 [2.533-3.256]   & 2.181 [1.964-2.668] \\
Persian            & 2.705 [2.575-2.884]   & 2.188 [2.029-2.398] \\
French (World)     & 2.831 [2.600-3.068]   & 2.068 [1.887-2.354] \\
French (Africa)    & 2.064 [1.945-2.204]   & 1.897 [1.774-2.061] \\
Hebrew             & 2.234 [1.928-2.543]   & 1.595 [1.405-1.867] \\
Hungarian          & 2.357 [2.125-2.710]   & 1.836 [1.604-2.160] \\
Norwegian          & 2.431 [2.071-3.287]   & 1.851 [1.574-2.203] \\
Polish             & 2.294 [2.067-2.589]   & 2.021 [1.734-2.359] \\
Romanian           & 2.300 [2.012-2.803]   & 1.783 [1.636-1.972] \\
Russian            & 2.651 [2.503-2.825]   & 2.050 [1.938-2.191] \\
Ukrainian          & 2.766 [2.410-3.292]   & 1.862 [1.666-2.106] \\
Chinese            & 3.068 [2.798-3.360]   & 2.406 [2.178-2.808] \\
\end{tabular}
\caption{\textbf{Average number of pageviews for men and women readers across surveys} We look at how many of the 16 surveys had a significant difference between men and women for the average number of pageviews in their reading session. Across all surveys, we see that men view significantly more articles per reading session than women do. Significance determined based on bootstrap resampling and 99\% confidence interval.}
\label{tab:data_session_pageviews_gender_language}
\end{table}

\begin{figure}[ht]
\centering
\includegraphics[width=1\textwidth]{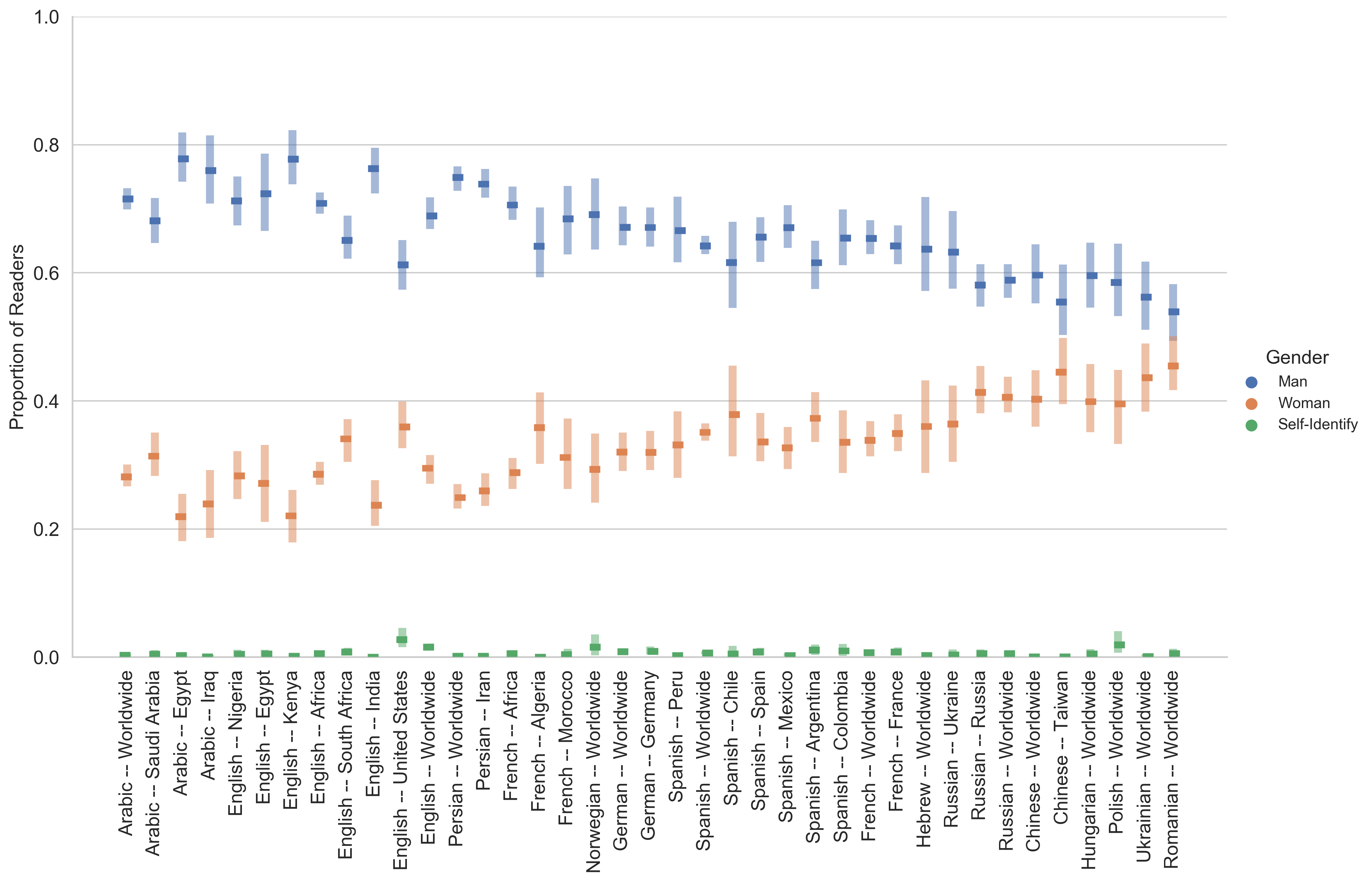}
\caption{\textbf{Proportion of Wikipedia readers by gender, country, and language} Results for readers who identify as men, women, or self-identify as otherwise. Substantial variation between countries within the same language can be seen---e.g., men comprise 61\% of readers in English Wikipedia in the United States but 76\% of readers in India. Only the country-language pairs for which we received 500 responses are shown. 99\% confidence intervals provided.}
\label{fig:reader_country_gender}
\end{figure}

\begin{table}[ht]
\centering
\begin{tabular}{l|c c c}
Information Need & Skew Sig. Men & Skew Sig. Women & No Sig. Diff. \\
\hline
In-Depth & 1 & 0 & 15 \\
Overview & 2 & 0 & 14 \\
Fact     & 0 & 7 & 9 \\
\end{tabular}
\caption{\textbf{Significant differences in age per information need across surveys.} We look at how many of the 16 surveys had a significant difference between men and women for their reported information need. For example, in 7 of the 16 surveys, women were more likely to report that they had come to Wikipedia looking for a fact than men (and there was no significant difference between men and women in the other 9 surveys). We see similar reported ``in-depth'' and ``overview'' information needs between men and women. Significance determined based on bootstrap resampling and 99\% confidence interval.}
\label{tab:data_gender_x_infoneed}
\end{table}

\begin{table}[ht]
\centering
\begin{tabular}{l|c c c}
Motivation & Skew Sig. Men & Skew Sig. Women & No Sig. Diff. \\
\hline
Bored / Random     & 12 & 0 & 4 \\
Current Event      & 7 & 0 & 9 \\
Work / School      & 0 & 7 & 9 \\
Intrinsic Learning & 4 & 1 & 11 \\
Media              & 1 & 2 & 13 \\
Conversation       & 1 & 1 & 14 \\
Personal Decision  & 2 & 0 & 14 \\
Other              & 1 & 1 & 14 \\
\end{tabular}
\caption{\textbf{Significant differences in age per motivation across surveys.} We look at how many of the 16 surveys had a significant difference between men and women for their reported motivation for why they were reading. For instance, we find that men reported being bored or searching at random significantly more often than women in 12 of 16 surveys (with no significant difference in the other 4). The only other consistent trends are that men reported looking up a current event significantly more often than  women in 7 of 16 surveys and women reported looking up information for work or school significantly more than men in 7 of 16 surveys. Significance determined based on bootstrap resampling and 99\% confidence interval.}
\label{tab:data_gender_x_motivation}
\end{table}

\begin{table}[ht]
\centering
\begin{tabular}{l|c c c}
Prior Knowledge & Skew Sig. Men & Skew Sig. Women & No Sig. Diff. \\
\hline
Familiar & 0 & 1 & 15 \\
Unfamiliar & 0 & 0 & 16 \\
\end{tabular}
\caption{\textbf{Significant differences in age per prior knowledge across surveys.} We look at how many of the 16 surveys had a significant difference between men and women for their reported prior knowledge of the article topic they were reading. For example, in 1 of the 16 surveys, women were significantly more likely than men to report being familiar with the topic of the article they were reading (and there was no significant difference between men and women in the other 15 surveys). We see no evidence of differing levels of familiarity between men and women. Significance determined based on bootstrap resampling and 99\% confidence interval.}
\label{tab:data_gender_x_familiarity}
\end{table}

\begin{table}[ht]
\centering
\begin{tabular}{l|c c c}
Age & Skew Sig. Men & Skew Sig. Women & No Sig. Diff. \\
\hline
18 to 24 & 0 & 6 & 10 \\
25 to 29 & 4 & 0 & 12 \\
30 to 39 & 7 & 0 & 9 \\
40 to 49 & 4 & 0 & 12 \\
50 to 59 & 4 & 0 & 12 \\
Over 60  & 3 & 2 & 11 \\
\end{tabular}
\caption{\textbf{Significant differences in age per gender across surveys.} We look at how many of the 16 surveys had a significant difference between men and women for how old they were. For example, in 6 of the 16 surveys, women were significantly more likely than men to report being age 18-24 (and there was no significant difference between men and women in the other 10 surveys). We see trends with varying levels of evidence that men are more likely to report being over the age of 24. Significance determined based on bootstrap resampling and 99\% confidence interval.}
\label{tab:data_gender_x_age}
\end{table}

\begin{table}[ht]
\centering
\begin{tabular}{l|c c c}
Educational Attainment & Skew Sig. Men & Skew Sig. Women & No Sig. Diff. \\
\hline
0 - 11 years  & 2 & 2 & 12 \\
12 years      & 2 & 3 & 11 \\
13 - 16 years & 0 & 1 & 15 \\
17 - 18 years & 0 & 0 & 16 \\
Over 18 years & 1 & 0 & 15 \\
\end{tabular}
\caption{\textbf{Significant differences in educational attainment per gender across surveys.} We look at how many of the 16 surveys had a significant difference between men and women for how many years of education they had completed. For example, in 1 of the 16 surveys, men were significantly more likely than women to report having attained more than 18 years of education (and there was no significant difference between men and women in the other 15 surveys). For most surveys, we see no significant differences, suggesting that differences in education level do not explain the observed differences in behavior between men and women. Significance determined based on bootstrap resampling and 99\% confidence interval.}
\label{tab:data_gender_x_education}

\end{table}

\begin{table}[ht]
\centering
\begin{tabular}{l|c c c}
Locale & Skew Sig. Men & Skew Sig. Women & No Sig. Diff. \\
\hline
City       & 1 & 2 & 13 \\
Suburbs    & 1 & 0 & 15 \\
Town       & 1 & 0 & 15 \\
Village    & 3 & 1 & 12 \\
Rural Area & 0 & 0 & 16 \\
\end{tabular}
\caption{\textbf{Significant differences in locale per gender across surveys.} We look at how many of the 16 surveys had a significant difference between men and women for the population density of the area in which they live. For example, in 1 of the 16 surveys, men were significantly more likely than women to report living in the suburbs (and there was no significant difference between men and women in the other 15 surveys). For most surveys, we see no significant difference. Significance determined based on bootstrap resampling and 99\% confidence interval.}
\label{tab:data_gender_x_locale}
\end{table}

\begin{table}[ht]
\centering
\begin{tabular}{l|c c c}
Language & Skew Sig. Men & Skew Sig. Women & No Sig. Diff. \\
\hline
Monolingual / Native  & 1 & 4 & 11 \\
Multilingual / Native & 0 & 3 & 13 \\
Non-native            & 4 & 0 & 12 \\
\end{tabular}
\caption{\textbf{Significant differences in native language per gender across surveys.} We look at how many of the 16 surveys had a significant difference between men and women for whether they were native speakers of the language in which they took the survey (and therefore the language version of Wikipedia that they were reading). For example, in 3 of the 16 surveys, women were significantly more likely than men to report being a native speaker of the survey language and at least one other language (and there was no significant difference between men and women in the other 13 surveys). We find limited evidence that women are more likely to be native speakers of just that language and men are more likely to be non-native speakers of the language that they are reading. Of note, 8 of these 12 significant differences come from English or French surveys (both worldwide and Africa-specific sampling). Significance determined based on bootstrap resampling and 99\% confidence interval.}
\label{tab:data_gender_x_language}
\end{table}

\begin{table}[ht]
\centering
\begin{tabular}{l|c c c}
Topic & Men & Women & Balanced \\
\hline
Culture---Sports & 12 & 0 & 4 \\
STEM---Technology & 12 & 0 & 4 \\
History and Society---Transportation & 11 & 0 & 5 \\
History and Society---Military and warfare & 10 & 0 & 6 \\
Culture---Broadcasting & 0 & 10 & 6 \\
STEM---Biology & 0 & 9 & 7 \\
STEM---Medicine & 0 & 9 & 7 \\
History and Society---Business and economics & 9 & 0 & 7 \\
Culture---Entertainment & 0 & 5 & 11 \\
Culture---Games and toys & 5 & 0 & 11 \\
History and Society---History and society & 1 & 4 & 11 \\
STEM---Physics & 4 & 0 & 12 \\
STEM---Space & 4 & 0 & 12 \\
Culture---Music & 0 & 3 & 13 \\
Culture---Performing arts & 0 & 3 & 13 \\
Geography---Africa & 3 & 0 & 13 \\
Geography---Americas & 2 & 1 & 13 \\
Geography---Asia & 1 & 2 & 13 \\
History and Society---Politics and government & 3 & 0 & 13 \\
Culture---Food and drink & 1 & 1 & 14 \\
Culture---Biography & 1 & 1 & 14 \\
Geography---Europe & 2 & 0 & 14 \\
STEM---Geosciences & 2 & 0 & 14 \\
STEM---Science & 2 & 0 & 14 \\
Culture---Plastic arts & 0 & 1 & 15 \\
Geography---Oceania & 0 & 1 & 15 \\
STEM---Chemistry & 0 & 1 & 15 \\
Culture---Internet culture & 0 & 0 & 16 \\
Culture---Philosophy and religion & 0 & 0 & 16 \\
Culture---Visual arts & 0 & 0 & 16 \\
History and Society---Education & 0 & 0 & 16 \\
STEM---Mathematics & 0 & 0 & 16 \\
\end{tabular}
\caption{\textbf{Significant differences in topics viewed per gender across surveys.} We look at how many of the 16 surveys had a significant difference between men and women for the topics that they viewed. For example, in 12 of the 16 survey populations, men were significantly more likely than women to read an article about sports (and there was no significant difference between men and women in the other 4 surveys). While men and women view most topics at the same rate, there are a few topics that men view more consistently (Sports, Technology, Military) or women view more consistently (Broadcasting, Biology, Medicine). Significance determined based on bootstrap resampling and 99\% confidence interval.}
\label{tab:topics}
\end{table}

\begin{table}[ht]
\centering
\tiny
\setlength\tabcolsep{2.5pt}
\begin{tabular}{l|*{16}{r}}
Variable & ar & zh & en (af) & en & fr (af) & fr & de & he & hu & no & pl & fa & ro & ru & es & uk \\
\hline
Intercept & $0.61***$ & $0.79***$ & $0.41***$ & $0.75***$ & $0.41*$ & $0.66***$ & $0.98***$ & $0.75$ & $0.88**$ & $0.29$ & $0.19$ & $1.06***$ & $0.77*$ & $-0.08$ & $0.34***$ & $-0.39$ \\
Age 25-29 & $0.69***$ & $0.53***$ & $0.54***$ & $0.20*$ & $0.72***$ & $-0.06$ & $0.20$ & $0.68$ & $0.07$ & $0.31$ & $-0.45$ & $0.11$ & $0.83**$ & $0.30**$ & $0.15*$ & $0.04$ \\
Age 30-39 & $0.65***$ & $0.37**$ & $0.58***$ & $0.38***$ & $0.79***$ & $0.18$ & $0.18$ & $-0.24$ & $-0.23$ & $-0.01$ & $0.40$ & $0.35***$ & $-0.10$ & $0.63***$ & $0.21**$ & $0.10$ \\
Age 40-49 & $0.83***$ & $0.47*$ & $0.49***$ & $-0.19$ & $1.02***$ & $-0.03$ & $0.07$ & $-0.22$ & $-0.24$ & $-0.12$ & $0.66$ & $0.60***$ & $-0.11$ & $0.37**$ & $0.15*$ & $0.53*$ \\
Age 50-59 & $0.87***$ & $0.15$ & $0.11$ & $-0.11$ & $1.34***$ & $0.18$ & $-0.28*$ & $-1.12*$ & $-0.65*$ & $0.16$ & $0.08$ & $0.97***$ & $0.66**$ & $0.15$ & $0.25**$ & $0.03$ \\
Age >60 & $1.31***$ & $1.08*$ & $0.40**$ & $-0.16$ & $2.05***$ & $-0.01$ & $-0.17$ & $-0.70$ & $-0.43*$ & $0.81**$ & $-1.60**$ & $0.70$ & $0.36$ & $0.00$ & $0.10$ & $0.51$ \\
Age NA & $-0.59***$ & $0.54$ & $-0.43*$ & $-0.53**$ & $0.26$ & $-0.42*$ & $-0.05$ & $-1.97***$ & $-1.39***$ & $-0.02$ & $0.06$ & $-0.26$ & $-0.11$ & $-0.08$ & $-0.42***$ & $-0.74$ \\
Topic: Culture.Arts & -- & -- & $-2.82$ & $13.19$ & -- & $0.18$ & $12.81$ & -- & -- & -- & -- & -- & -- & -- & $0.80$ & -- \\
Topic: Culture.Broadcasting & $-0.58**$ & $-0.57*$ & $-0.54***$ & $-0.67***$ & $-0.48$ & $-0.51**$ & $-0.42$ & $-0.65$ & $-1.20**$ & $-0.03$ & $-16.55$ & $-0.29$ & $-0.09$ & $-0.53*$ & $-0.58***$ & $-2.28*$ \\
Topic: Culture.Crafts/Hobbies & $12.97$ & -- & $2.65$ & $0.00$ & $-1.09$ & $13.25$ & $3.23$ & -- & -- & $16.05$ & -- & $-1.40$ & -- & $13.94$ & $-1.45$ & $-0.51$ \\
Topic: Culture.Entertainment & $-0.53**$ & $-0.18$ & $-0.07$ & $-0.19$ & $-0.06$ & $0.17$ & $-0.37$ & $0.44$ & $-0.97**$ & $0.57$ & $0.18$ & $-0.64***$ & $1.10$ & $-0.19$ & $-0.12$ & $-0.07$ \\
Topic: Culture.Food/drink & $-0.94***$ & $1.89***$ & $-0.79***$ & $0.31$ & $-0.19$ & $-0.12$ & $-0.29$ & $4.39$ & $0.17$ & $-0.35$ & $-0.23$ & $0.47$ & $0.29$ & $-0.09$ & $-0.02$ & $-0.22$ \\
Topic: Culture.Games/Toys & $1.02*$ & $0.74$ & $0.50$ & $1.28***$ & $0.72$ & $0.57$ & $-0.68$ & $16.53$ & $15.30$ & $16.37$ & $0.91$ & $0.42$ & $0.12$ & $1.45***$ & $1.03***$ & $2.02*$ \\
Topic: Culture.Internet culture & $0.76$ & $13.33$ & $12.74$ & $0.00$ & $-2.11$ & $1.75$ & $-2.45$ & $15.62$ & $15.33$ & -- & -- & $0.79$ & $-0.79$ & $-2.73$ & $-0.11$ & $15.20$ \\
Topic: Culture.Media & -- & -- & $12.32$ & $-4.48$ & $12.70$ & $13.10$ & -- & -- & -- & -- & -- & -- & -- & $13.60$ & -- & $-15.17$ \\
Topic: Culture.Music & $-0.27$ & $-1.68**$ & $0.25$ & $0.62**$ & $1.81$ & $0.24$ & $0.36$ & $1.47$ & $0.16$ & $16.88$ & $1.27$ & $0.00$ & $0.68$ & $0.41$ & $0.05$ & $0.11$ \\
Topic: Culture.Performing arts & $-1.46$ & $15.01$ & $-1.95**$ & $-0.82$ & $-4.20*$ & $-0.97$ & $-0.33$ & -- & $-5.35$ & -- & $-16.80$ & $-2.29*$ & $-0.85$ & $-1.85$ & $0.04$ & $0.86$ \\
Topic: Culture.Philosophy/Religion & $-0.46***$ & $-0.72$ & $0.12$ & $-0.23$ & $-0.31$ & $-0.09$ & $0.49$ & $-0.61$ & $1.25*$ & $3.76$ & $0.28$ & $-0.18$ & $-0.01$ & $0.11$ & $-0.11$ & $0.51$ \\
Topic: Culture.Plastic arts & $1.97$ & $15.07$ & $1.63$ & $-0.06$ & $1.31$ & $1.60$ & $1.39$ & $-16.34$ & $0.12$ & $-1.97$ & $15.57$ & $0.04$ & -- & $-1.19$ & $-0.64$ & $-0.01$ \\
Topic: Culture.Sports & $0.91***$ & $2.46*$ & $0.73***$ & $0.98***$ & $1.83***$ & $0.67**$ & $0.93**$ & $15.25$ & $0.23$ & $0.78$ & $0.21$ & $0.19$ & $0.68$ & $0.85**$ & $1.07***$ & $0.41$ \\
Topic: Culture.Visual arts & $-0.61$ & $0.25$ & $0.21$ & $0.19$ & $0.62$ & $0.31$ & $1.75$ & $-3.49$ & $1.61$ & $1.58$ & $15.92$ & $-0.11$ & $1.19$ & $0.05$ & $0.36$ & $-1.55$ \\
Topic: Geo.Africa & $0.43**$ & $-0.23$ & $0.21$ & $-0.46$ & $-0.26$ & $0.21$ & $-1.04*$ & $18.67$ & $-1.19$ & $-2.67$ & $16.36$ & $0.39$ & $-3.36$ & $-0.40$ & $1.28**$ & $3.71$ \\
Topic: Geo.Americas & $0.58*$ & $2.84*$ & $0.29$ & $0.21$ & $0.43$ & $0.27$ & $-0.02$ & $15.24$ & $1.48$ & $2.26$ & $-0.04$ & $0.75*$ & $0.54$ & $0.21$ & $0.04$ & $0.55$ \\
Topic: Geo.Antarctica & -- & -- & -- & $-2.71$ & -- & -- & -- & -- & -- & -- & -- & -- & -- & $-14.54$ & $11.91$ & -- \\
Topic: Geo.Asia & $0.21$ & $-0.17$ & $-0.20$ & $0.04$ & $-0.04$ & $-0.34$ & $0.65$ & $0.17$ & $-0.01$ & $0.32$ & $0.84$ & $-0.06$ & $-0.47$ & $-0.39**$ & $0.05$ & $-0.76$ \\
Topic: Geo.Bodies of water & $12.12$ & $14.07$ & $-0.87$ & $13.27$ & $12.90$ & $12.24$ & $12.46$ & -- & $-15.28$ & -- & -- & $10.76$ & $1.93$ & $14.07$ & $-0.44$ & $14.82$ \\
Topic: Geo.Europe & $0.04$ & $0.89$ & $0.38*$ & $0.00$ & $0.20$ & $0.16$ & $0.21$ & $0.10$ & $0.50*$ & $0.49$ & $0.29$ & $-0.06$ & $0.04$ & $0.28*$ & $-0.11$ & $-0.04$ \\
Topic: Geo.Landforms & $-0.69$ & $13.96$ & $13.02$ & $0.06$ & $2.04$ & $12.81$ & $0.17$ & -- & -- & -- & -- & $-0.81$ & $0.73$ & $14.30$ & $11.82$ & -- \\
Topic: Geo.Maps & $11.89$ & -- & $11.71$ & $13.38$ & $-14.25$ & $12.37$ & $13.21$ & -- & -- & -- & -- & -- & -- & $14.23$ & $12.20$ & -- \\
Topic: Geo.Oceania & $-1.51$ & $-0.12$ & $-0.58$ & $0.61$ & $-15.95$ & $-0.86$ & $0.01$ & -- & $3.82$ & $1.12$ & -- & $-0.83$ & $-0.02$ & $-1.32$ & $-0.15$ & $-15.77$ \\
Topic: Geo.Parks & $-0.96$ & $14.31$ & $-1.41$ & $-0.74$ & $-2.40$ & $12.66$ & -- & $16.89$ & -- & $15.52$ & -- & $11.17$ & $-0.01$ & $14.83$ & $0.49$ & $1.55$ \\
Topic: HisSoc.Business/economics & $0.06$ & $2.33**$ & $0.52**$ & $0.47*$ & $0.10$ & $0.63*$ & $1.07***$ & $0.27$ & $-0.37$ & $2.77$ & $2.23*$ & $0.79*$ & $-2.07**$ & $0.99***$ & $0.35*$ & $0.65$ \\
Topic: HisSoc.Education & $-0.48$ & $0.23$ & $-0.58*$ & $-1.10**$ & $-0.29$ & $-0.43$ & $0.61$ & $15.86$ & $-3.93$ & -- & $15.12$ & $-0.25$ & $-2.40$ & $0.73$ & $-0.11$ & $-0.11$ \\
Topic: HisSoc.History/society & $-0.01$ & $0.53$ & $0.06$ & $-0.18$ & $0.34$ & $-0.29$ & $0.21$ & $0.25$ & $0.60$ & $-0.22$ & $-0.12$ & $-0.26$ & $0.06$ & $-0.45*$ & $-0.02$ & $-0.87$ \\
Topic: HisSoc.Military/warfare & $0.57$ & $3.41*$ & $1.57***$ & $3.71***$ & $-0.42$ & $0.96*$ & $0.29$ & $15.53$ & $1.69$ & $15.56$ & $16.75$ & $2.31**$ & $1.79$ & $1.58***$ & $0.34$ & $15.30$ \\
Topic: HisSoc.Politics/government & $0.05$ & $-0.25$ & $-0.10$ & $0.24$ & $-0.51$ & $-0.14$ & $-0.44$ & $-1.01$ & $-0.32$ & $-0.35$ & $-0.86$ & $0.72*$ & $-0.20$ & $0.26$ & $0.17$ & $0.80$ \\
Topic: HisSoc.Transportation & $0.57$ & $1.28$ & $1.23***$ & $1.11***$ & $0.20$ & $0.33$ & $1.23***$ & $0.74$ & $1.51**$ & $1.56*$ & $1.76*$ & $1.48**$ & $-0.57$ & $1.47***$ & $1.00**$ & $0.60$ \\
Topic: STEM.Biology & $-0.49**$ & $0.60*$ & $-0.21$ & $-0.07$ & $-0.44*$ & $-0.26$ & $-0.47**$ & $-0.14$ & $0.51*$ & $0.00$ & $0.19$ & $-0.20$ & $-1.23***$ & $-0.37*$ & $-0.26**$ & $-0.42$ \\
Topic: STEM.Chemistry & $-0.02$ & $0.22$ & $0.75*$ & $0.13$ & $-0.61$ & $-0.33$ & $0.77*$ & $-0.63$ & $-1.27$ & $0.93$ & $0.25$ & $-0.38$ & $-1.13$ & $0.77$ & $0.51*$ & $15.85$ \\
Topic: STEM.Engineering & $-1.46$ & $14.36$ & $1.55$ & $-0.99$ & $1.11$ & $13.48$ & $13.00$ & -- & $-15.84$ & -- & -- & $-1.68$ & $-0.14$ & $0.07$ & $0.27$ & -- \\
Topic: STEM.Geosciences & $0.33$ & $-4.25$ & $1.70*$ & $0.94$ & $0.15$ & $0.01$ & $0.77$ & $-16.64$ & $0.10$ & $15.56$ & $-16.41$ & $0.84$ & $-2.72$ & $0.31$ & $0.34$ & $1.46$ \\
Topic: STEM.Information science & -- & -- & $12.74$ & $13.73$ & -- & $13.09$ & $12.43$ & -- & -- & -- & -- & -- & -- & -- & -- & -- \\
Topic: STEM.Mathematics & $12.35$ & $5.03$ & $-0.11$ & $14.06$ & $-0.21$ & $-0.34$ & $0.94$ & $14.98$ & $-0.69$ & -- & $16.52$ & $-1.60***$ & $-0.77$ & $2.08$ & $0.56$ & $-16.13$ \\
Topic: STEM.Medicine & $-0.64***$ & $-0.10$ & $-0.68***$ & $-0.57***$ & $-0.97***$ & $-0.28$ & $-0.76***$ & $-1.20**$ & $-0.44$ & $-0.14$ & $-0.44$ & $-0.72***$ & $-0.69*$ & $-0.77***$ & $-0.34***$ & $-0.45$ \\
Topic: STEM.Meteorology & -- & $-15.05$ & -- & $1.03$ & $-1.24$ & $3.02$ & -- & -- & -- & -- & -- & $-13.24$ & $12.19$ & $-2.28$ & $0.76$ & -- \\
Topic: STEM.Physics & $0.96$ & $-1.87$ & $1.11$ & $2.63$ & $1.46$ & $-0.40$ & $0.52$ & $7.17$ & $15.32$ & $16.21$ & $6.26$ & $2.04$ & $-0.24$ & $14.46$ & $0.71*$ & $5.85$ \\
Topic: STEM.Science & $-0.50$ & $1.47$ & $0.70$ & $3.24$ & $-0.94$ & $2.49*$ & $0.28$ & $-16.64$ & $15.85$ & $17.61$ & $1.10$ & $-1.35$ & $-1.87$ & $2.32*$ & $0.67$ & $-2.28$ \\
Topic: STEM.Space & $0.02$ & $-0.64$ & $0.48$ & $1.18*$ & $13.64$ & $0.68$ & $1.22$ & $15.06$ & $1.27$ & $-1.53$ & $0.12$ & $0.10$ & $-0.35$ & $1.26*$ & $-0.02$ & $2.59$ \\
Topic: STEM.Technology & $0.27$ & $1.51***$ & $0.60***$ & $0.41*$ & $0.24$ & $0.96***$ & $0.72**$ & $1.20$ & $1.85**$ & $-0.08$ & $6.73$ & $0.97***$ & $1.63**$ & $1.07***$ & $0.80***$ & $0.94*$ \\
Topic: STEM.Time & $-0.16$ & $-1.90$ & $-0.65$ & $-2.36$ & $0.31$ & $1.99$ & $1.21$ & $-1.64$ & $3.45$ & $1.03$ & $16.49$ & $0.86$ & $1.11$ & $0.07$ & $-0.41$ & $0.08$ \\
No Topic & $0.03$ & $0.26$ & $0.14$ & $0.19$ & $-0.12$ & $0.09$ & $0.21$ & $0.37$ & $0.40$ & $0.42$ & $0.16$ & $-0.04$ & $-0.11$ & $0.16$ & $0.18*$ & $0.18$ \\
Algeria & -- & -- & -- & -- & $-0.53***$ & -- & -- & -- & -- & -- & -- & -- & -- & -- & -- & -- \\
Argentina & -- & -- & -- & -- & -- & -- & -- & -- & -- & -- & -- & -- & -- & -- & $-0.02$ & -- \\
Chile & -- & -- & -- & -- & -- & -- & -- & -- & -- & -- & -- & -- & -- & -- & $-0.05$ & -- \\
Colombia & -- & -- & -- & -- & -- & -- & -- & -- & -- & -- & -- & -- & -- & -- & $0.12$ & -- \\
Egypt & $0.35***$ & -- & $0.10$ & -- & -- & -- & -- & -- & -- & -- & -- & -- & -- & -- & -- & -- \\
France & -- & -- & -- & -- & -- & $-0.23**$ & -- & -- & -- & -- & -- & -- & -- & -- & -- & -- \\
Germany & -- & -- & -- & -- & -- & -- & $-0.05$ & -- & -- & -- & -- & -- & -- & -- & -- & -- \\
Hungary & -- & -- & -- & -- & -- & -- & -- & -- & $-0.33$ & -- & -- & -- & -- & -- & -- & -- \\
India & -- & -- & -- & $0.22*$ & -- & -- & -- & -- & -- & -- & -- & -- & -- & -- & -- & -- \\
Iran & -- & -- & -- & -- & -- & -- & -- & -- & -- & -- & -- & $-0.26***$ & -- & -- & -- & -- \\
Iraq & $0.38**$ & -- & -- & -- & -- & -- & -- & -- & -- & -- & -- & -- & -- & -- & -- & -- \\
Israel & -- & -- & -- & -- & -- & -- & -- & $-0.61$ & -- & -- & -- & -- & -- & -- & -- & -- \\
Kenya & -- & -- & $0.30**$ & -- & -- & -- & -- & -- & -- & -- & -- & -- & -- & -- & -- & -- \\
Mexico & -- & -- & -- & -- & -- & -- & -- & -- & -- & -- & -- & -- & -- & -- & $0.16*$ & -- \\
Morocco & -- & -- & -- & -- & $-0.26*$ & -- & -- & -- & -- & -- & -- & -- & -- & -- & -- & -- \\
Nigeria & -- & -- & $-0.04$ & -- & -- & -- & -- & -- & -- & -- & -- & -- & -- & -- & -- & -- \\
Norway & -- & -- & -- & -- & -- & -- & -- & -- & -- & $0.10$ & -- & -- & -- & -- & -- & -- \\
Peru & -- & -- & -- & -- & -- & -- & -- & -- & -- & -- & -- & -- & -- & -- & $0.26**$ & -- \\
Poland & -- & -- & -- & -- & -- & -- & -- & -- & -- & -- & $-0.18$ & -- & -- & -- & -- & -- \\
Romania & -- & -- & -- & -- & -- & -- & -- & -- & -- & -- & -- & -- & $0.14$ & -- & -- & -- \\
Russia & -- & -- & -- & -- & -- & -- & -- & -- & -- & -- & -- & -- & -- & $0.04$ & -- & -- \\
Saudi Arabia & $-0.08$ & -- & -- & -- & -- & -- & -- & -- & -- & -- & -- & -- & -- & -- & -- & -- \\
South Africa & -- & -- & $-0.31***$ & -- & -- & -- & -- & -- & -- & -- & -- & -- & -- & -- & -- & -- \\
Spain & -- & -- & -- & -- & -- & -- & -- & -- & -- & -- & -- & -- & -- & -- & $0.12$ & -- \\
Taiwan & -- & $-0.35**$ & -- & -- & -- & -- & -- & -- & -- & -- & -- & -- & -- & -- & -- & -- \\
Ukraine & -- & -- & -- & -- & -- & -- & -- & -- & -- & -- & -- & -- & -- & $0.22*$ & -- & $-0.20$ \\
United States & -- & -- & -- & $-0.32***$ & -- & -- & -- & -- & -- & -- & -- & -- & -- & -- & -- & -- \\
Edu: 12 years & $-0.10$ & $-0.65**$ & $0.18$ & $-0.06$ & $0.43*$ & $-0.02$ & $-0.46***$ & $-0.26$ & $-0.56**$ & $0.35$ & $-0.65*$ & $-0.39***$ & $-0.58**$ & $0.45***$ & $0.18*$ & $1.34***$ \\
Edu: 13-16 years & $-0.36***$ & $-0.44*$ & $0.03$ & $-0.07$ & $0.23$ & $-0.04$ & $-0.33***$ & $-0.56$ & $-0.06$ & $0.33$ & $0.02$ & $-0.19$ & $-0.73***$ & $-0.09$ & $0.08$ & $0.23$ \\
Edu: 17-18 years & $-0.43***$ & $-0.31$ & $0.00$ & $-0.05$ & $0.05$ & $0.14$ & $-0.24$ & $0.51$ & $0.18$ & $-0.18$ & $0.06$ & $-0.41***$ & $-0.20$ & $-0.38**$ & $0.07$ & $0.33$ \\
Edu: >19 years & $-0.29$ & $0.10$ & $-0.02$ & $-0.03$ & $0.12$ & $0.25*$ & $-0.05$ & $0.05$ & $0.07$ & $-0.11$ & $0.41$ & $-0.25$ & $-0.70**$ & $0.08$ & $0.11$ & $0.50$ \\
Edu: NA & $-0.54***$ & $-0.98**$ & $-0.13$ & $-0.16$ & $0.08$ & $0.28*$ & $-0.47$ & $3.49$ & $-0.08$ & $0.20$ & $-0.06$ & $0.23$ & $-0.79*$ & $0.19$ & $-0.09$ & $0.77*$ \\
Info Depth: In-depth & $0.27***$ & $0.05$ & $0.09$ & $0.12$ & $0.20$ & $0.44***$ & $0.16$ & $0.91*$ & $-0.17$ & $0.32$ & $0.24$ & $0.36***$ & $-0.23$ & $0.18*$ & $0.18***$ & $0.42*$ \\
Info Depth: Other & $-0.34$ & $-0.94*$ & $0.35$ & $-0.26$ & $1.17*$ & $-0.14$ & $-0.02$ & $0.70$ & $0.47$ & $-2.24**$ & $0.27$ & $0.55$ & $0.16$ & $0.43$ & $-0.23$ & $0.83$ \\
Info Depth: Overview & $0.32***$ & $0.11$ & $0.10$ & $0.08$ & $0.25*$ & $0.30***$ & $0.04$ & $0.56$ & $0.09$ & $0.08$ & $0.12$ & $0.47***$ & $-0.30$ & $0.22**$ & $0.20***$ & $0.62***$ \\
Lang: Native+ & $-0.02$ & $-0.23*$ & $0.01$ & $0.05$ & $-0.32*$ & $-0.37***$ & $-0.11$ & $0.15$ & $-0.06$ & $-0.42$ & $-0.47$ & $0.12$ & $0.07$ & $0.10$ & $0.00$ & $-0.10$ \\
Lang: Non-native & $0.22$ & $-0.11$ & $0.34***$ & $0.27**$ & $0.06$ & $-0.35*$ & $-0.31*$ & $-0.21$ & $-0.17$ & $-0.63$ & $0.44$ & $0.60***$ & $-0.04$ & $0.28*$ & $0.15$ & $0.35$ \\
Lang: NA & $0.92$ & $-1.34$ & $0.20$ & $-0.29$ & $0.76*$ & $0.01$ & $-1.15$ & $16.52$ & $0.64$ & $0.36$ & $-0.33$ & $1.01*$ & $-0.15$ & $0.34$ & -- & $2.68$ \\
Locale: NA & $0.07$ & $-0.48$ & $-0.19$ & $-0.62**$ & $-0.02$ & $-0.17$ & $-0.12$ & $-1.02$ & $-0.05$ & $-1.77$ & $0.45$ & $-0.05$ & $0.55$ & $-0.58**$ & $-0.72***$ & $0.66$ \\
Locale: rural & $0.14$ & $0.11$ & $0.13$ & $-0.27$ & $0.36$ & $-0.06$ & $-0.32$ & $-3.69$ & $-0.68$ & $-0.16$ & $-6.76$ & $0.03$ & $-0.28$ & $0.48$ & $-0.33**$ & $1.55*$ \\
Locale: suburbs & $0.02$ & $-0.24$ & $0.01$ & $0.21*$ & $0.26$ & $-0.09$ & $0.07$ & $0.32$ & $0.38$ & $0.04$ & $0.31$ & $0.13$ & $-0.35$ & $0.04$ & $-0.02$ & $-0.29$ \\
Locale: town & $0.18**$ & $0.03$ & $-0.02$ & $-0.14$ & $0.22*$ & $-0.15$ & $0.01$ & $0.74*$ & $-0.15$ & $-0.10$ & $0.32$ & $0.16*$ & $0.36*$ & $-0.02$ & $-0.05$ & $0.07$ \\
Locale: village & $0.56***$ & $-0.10$ & $0.25*$ & $-0.19$ & $0.28$ & $-0.05$ & $-0.10$ & $0.55$ & $0.01$ & $-0.09$ & $0.16$ & $0.61*$ & $-0.37*$ & $0.07$ & $-0.12$ & $0.34$ \\
\hline
Sample Size & 7380 & 2019 & 7711 & 5849 & 2953 & 4208 & 3959 & 560 & 1188 & 710 & 641 & 6682 & 1259 & 4404 & 11277 & 1107 \\
Residual Degrees of Freedom & 7315 & 1957 & 7642 & 5780 & 2887 & 4141 & 3895 & 504 & 1129 & 655 & 586 & 6619 & 1198 & 4337 & 11207 & 1046 \\ 
\end{tabular}
\caption{\label{tab:regression}Significant predictors of men readers based on individual logistic regression models for each survey. Note, this reduces gender to a binary, but we had insufficient data from individuals with non-binary gender identities to run more appropriate analyses. Each survey is a separate column, as indicated by its language code---e.g., ``fa'' is ``Persian'', ``en (af)'' indicates ``English (Africa-only)''. Each row is an independent variable in a logistic regression. Most variables appear in all models (like age) but some (like specific countries) only are used for certain surveys' models. ``NA'' as in ``Age: NA'' indicates the respondent selected ``Prefer not to say''. Reference levels same for all models and as follows: Age 18-24, Topic: Biographies, Country: Other, Edu: <12 years, Info. Depth: Fact, Language: Native only, Locale: city. Significance codes:  *** $<0.001$; ** $<0.01$
}
\end{table}

\begin{table}[ht]
\centering
\begin{tabular}{l l l| l l l}
Rank & Wikidata Item & English Title  & \% Men & \% Women & Total Views across All Surveys \\
\hline
1 & Q486 & Chernobyl disaster & $0.682$ & $0.282$ & $305$ \\
2 & Q11831154 & 2019 Africa Cup of Nations & $0.852$ & $0.107$ & $243$ \\
3 & Q19771 & G20 & $0.846$ & $0.123$ & $228$ \\
4 & Q4630358 & 2019 Copa América & $0.847$ & $0.106$ & $189$ \\
5 & Q517906 & Gabriel Calderón & $0.951$ & $0.038$ & $185$ \\
6 & Q52442498 & Mahnoosh Sadeghi & $0.588$ & $0.362$ & $177$ \\
7 & Q4630361 & 2019 FIFA Women's World Cup & $0.829$ & $0.134$ & $164$ \\
8 & Q3113912 & Mehdi Hashemi & $0.510$ & $0.432$ & $155$ \\
9 & Q30 & United States & $0.785$ & $0.161$ & $149$ \\
10 & Q59784750 & Deaths in 2019 & $0.824$ & $0.128$ & $148$ \\
11 & Q23781155 & Avengers: Endgame & $0.844$ & $0.099$ & $141$ \\
12 & Q27985819 & Spider-Man: Far From Home & $0.791$ & $0.171$ & $129$ \\
13 & Q48741246 & Chernobyl (miniseries) & $0.697$ & $0.287$ & $122$ \\
14 & Q550653 & Max Wright & $0.793$ & $0.182$ & $121$ \\
15 & Q392108 & List of highest-grossing films & $0.825$ & $0.149$ & $114$ \\
16 & Q260725 & Megan Rapinoe & $0.723$ & $0.241$ & $112$ \\
17 & Q3109977 & Golab Adineh & $0.519$ & $0.387$ & $106$ \\
18 & Q29564107 & Billie Eilish & $0.629$ & $0.343$ & $105$ \\
19 & Q43416 & Keanu Reeves & $0.705$ & $0.257$ & $105$ \\
20 & Q91772 & Lisa Martinek & $0.755$ & $0.235$ & $102$ \\
21 & Q178750 & Copa América & $0.822$ & $0.139$ & $101$ \\
22 & Q1386100 & Ezzat Abou Aouf & $0.694$ & $0.286$ & $98$ \\
23 & Q52 & Wikipedia & $0.821$ & $0.147$ & $95$ \\
24 & Q233510 & Alex Morgan & $0.791$ & $0.187$ & $91$ \\
25 & Q19323 & FIFA Women's World Cup & $0.813$ & $0.132$ & $91$ \\
26 & Q22686 & Donald trump & $0.818$ & $0.159$ & $88$ \\
27 & Q794 & Iran & $0.793$ & $0.161$ & $87$ \\
28 & Q20650884 & 2019 CONCACAF Gold Cup & $0.837$ & $0.116$ & $86$ \\
29 & Q159 & Russland & $0.729$ & $0.235$ & $85$ \\
30 & Q9682 & Elizabeth II & $0.634$ & $0.280$ & $82$ \\
31 & Q28443710 & Dark (TV series) & $0.695$ & $0.293$ & $82$ \\
32 & Q362 & World War II & $0.716$ & $0.210$ & $81$ \\
33 & Q56364616 & -- & $0.556$ & $0.420$ & $81$ \\
34 & Q27733841 & Bob Collymore & $0.725$ & $0.250$ & $80$ \\
35 & Q83145 & Africa Cup of Nations & $0.855$ & $0.105$ & $76$ \\
36 & Q567 & Angela Merkel & $0.773$ & $0.200$ & $75$ \\
37 & Q183 & Germany & $0.824$ & $0.108$ & $74$ \\
38 & Q352 & Adolf Hitler & $0.757$ & $0.189$ & $74$ \\
39 & Q1368046 & Solar eclipse of July 2, 2019 & $0.704$ & $0.282$ & $71$ \\
40 & Q8646 & Hong Kong & $0.814$ & $0.100$ & $70$ \\
41 & Q18517638 & Toy Story 4 & $0.714$ & $0.243$ & $70$ \\
42 & Q11571 & Cristiano Ronaldo & $0.754$ & $0.203$ & $69$ \\
43 & Q148 & China & $0.855$ & $0.087$ & $69$ \\
44 & Q131723 & Bitcoin & $0.812$ & $0.159$ & $69$ \\
45 & Q145 & United Kingdom & $0.750$ & $0.191$ & $68$ \\
46 & Q33038474 & 2019 G20 Osaka summit & $0.809$ & $0.132$ & $68$ \\
47 & Q5873 & Sexual intercourse & $0.701$ & $0.269$ & $67$ \\
48 & Q2831 & Michael Jackson & $0.727$ & $0.227$ & $66$ \\
49 & Q361 & World War I & $0.769$ & $0.215$ & $65$ \\
50 & Q20109415 & Noora Hashemi & $0.531$ & $0.375$ & $64$ \\
\end{tabular}
\caption{\textbf{Top-50 most read articles by survey respondents.} Articles most-viewed by survey respondents as well as what proportion of the views came from respondents who identified as men or respondents who identified as women. The remaining pageviews came from respondents who did not provide a gender or self-identified. Views were aggregated by Wikidata item and English article title given for interpretability where available but many of the views came from Wikipedia language editions that were not English. 
}
\label{tab:most_popular_articles}
\end{table}

\end{document}